# Unveiling Lattice Dynamics and a Hidden Structural Transition in the 2D Ferromagnet $AgVP_2Se_6$ via Raman Spectroscopy

Thi Hai Yen Pham[1,2*], Po-hao Chang[1,2], Aidan C. Malloy[1,2], Seng Huat Lee[3], Zhiqiang Mao[3], Maria F. Munoz[4], Riccardo Torsi[4], Angela R. Hight Walker[4], Igor Mazin[1,2], Patrick M. Vora[1,2*]

[1]Department of Physics and Astronomy, George Mason University, Fairfax, VA 22030

[2]Quantum Science and Engineering Center, George Mason University, Fairfax, VA 22030

[3]2D Crystal Consortium, Materials Research Institute, The Pennsylvania State University, University Park, PA 16802

[4]Quantum Measurement Division, Physical Measurement Laboratory, NIST, Gaithersburg, MD 20899

[*]Correspondence: tpham62@gmu.edu, pvora@gmu.edu

**Abstract**

Transition-metal chalcophosphates (TMCs) are two-dimensional (2D) van der Waals materials supporting a broad range of electronic and magnetic properties. Among quaternary TMCs, $AgVP_2Se_6$ is a rare ferromagnet (FM) with a triangular V sublattice, in contrast to the antiferromagnetic (AFM) zigzag chains of the sulfide $AgVP_2S_6$. Prior Raman studies of $AgVP_2Se_6$ have been performed far above the Curie temperature ($T_C$), preventing studies of spin-phonon coupling. Here, we report polarization-resolved, temperature-dependent Raman and magneto-Raman spectroscopy of $AgVP_2Se_6$ single crystals grown by chemical vapor transport (CVT) and flux methods, across the FM transition, complemented by density functional theory (DFT) calculations of phonon, optical and magnetic properties. Cryogenic Raman spectra resolve up to thirty peaks, while angle-resolved measurements enable their symmetry assignment. Spectral differences between CVT- and flux-grown crystals are traced back to the presence of different interlayer stacking domains in these samples. Temperature- and field-dependent Raman spectra remain largely unchanged across $T_C$, suggesting weak spin-phonon coupling. However, pronounced spectral alterations near 100 K suggest a potential structural phase transition. Our DFT calculations further reconcile the large discrepancy between the transport activation gap (≈ 0.325 eV) and optical absorption edge (≈ 2.14 eV), attributing the former to a transition to the V-*d* upper Hubbard band that is optically dark but thermally accessible. We also suggest a theoretical explanation of the qualitative difference (reproduced by DFT calculations) between the magnetic properties of the sulfide and selenide. These results provide key information regarding the lattice dynamics of $AgVP_2Se_6$ and will guide future applications in spin-based electronics.

Transition metal chalcophosphates (TMCs) are a versatile family of layered two-dimensional (2D) materials with the general formula $MPX_3$, where M = 3*d* metal and X = S or Se. The chemical flexibility of these compounds leads to tunable optical bandgaps (≈ 1.3 eV to 3.5 eV) as well as magnetic and ferroelectric behaviors.[1] When magnetic atoms occupy the metal sublattice, as in $MnPSe_3$[2] and $FePS_3$,[3] TMCs exhibit long-range magnetic order even in the monolayer limit, making them a compelling platform for studies of low-dimensional magnetism and spintronics.[1] Most ternary TMCs order antiferromagnetically (AFM), with Néel temperature of 80 K to 150 K.[1–5] In quaternary TMCs ($MM'P_2X_6$ where M and M' are transition metals), two different cations share the metal sublattice, one nonmagnetic (*e.g.,* $Ag^+$, $Cu^+$) and one magnetic (*e.g.,* $V^{3+}$, $Cr^{3+}$), offering an additional degree of freedom for tailoring magnetic and ferroic properties not accessible in their ternary counterparts.[6] Replacing every other magnetic ion with a nonmagnetic species reduces the density of magnetic sites in the lattice. Magnetic order is suppressed to lower temperatures and the ground state can be either ferromagnetic (FM)[7–10] or antiferromagnetic (AFM).[6,11–14]

How a quaternary TMC orders FM or AFM depends on how the magnetic cations arrange themselves. It was noted in the literature that the triangular arrangement yields in-plane FM order, while the zigzag one generates AFM, but no microscopic explanation has been presented so far.[1,6,7,10,11,15] Our DFT calculations reproduce this trend and indicate that when transition metals appear as nearest neighbors on the honeycomb lattice, as in the ternary $MPX_3$, a classical superexchange interaction appears that strongly favors an AFM order.[16] This still holds in the zigzag quaternaries. However, in the triangular arrangement no magnetic sites are nearest neighbors. There is a transferred interaction due to polarization of the intermediate ion (Ag, in this case); after integrating out the bridging atom's states, a (much weaker) net FM appears between the transition metal sites. This is similar to the much-better studied case of FM double perovskites.[17]

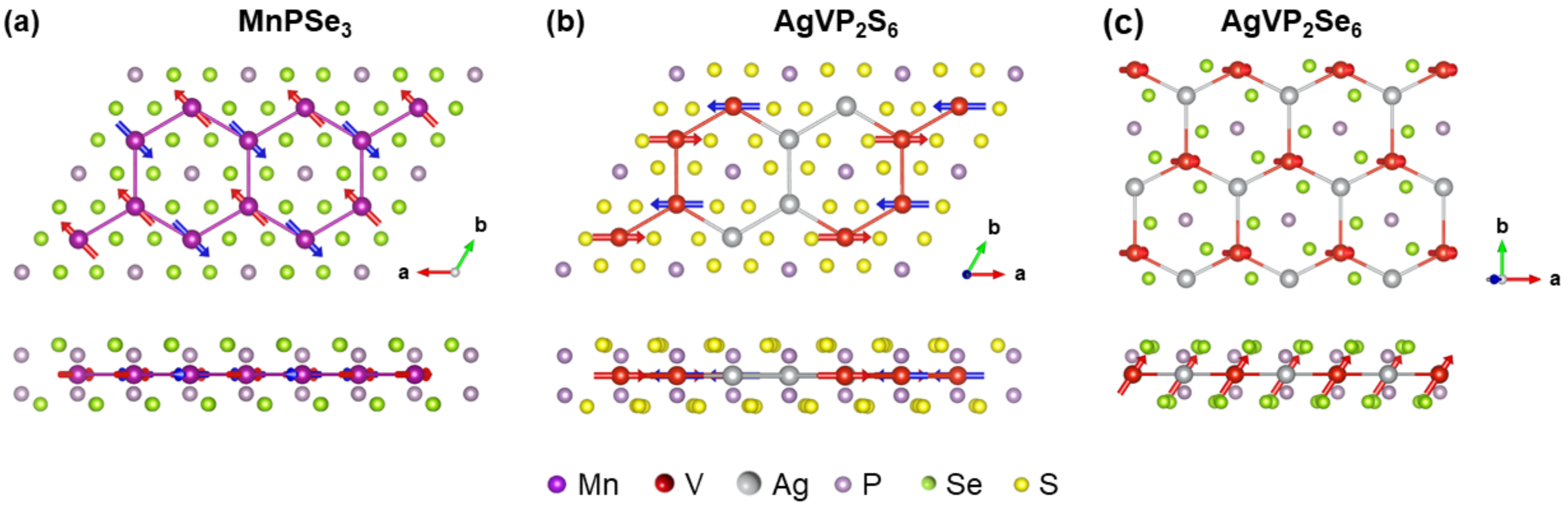

**Figure 1** – Magnetic orders in the ternary and quaternary TMC compounds with top (upper) and side (lower) views of their monolayers. (a) Crystal and magnetic structure of the Mn plane in $MnPSe_3$ with the strong nearest Mn AFM bonds illustrated. (b) Crystal structure of $AgVP_2S_6$, where the symmetry is now broken and V atoms form continuous AFM zigzag chains. (c) Crystal structure of $AgVP_2Se_6$ shows that the three-fold symmetry is preserved, and no V atoms are nearest neighbors, resulting in an overall FM order. The Mn, V, Ag, P, Se, and S atoms are depicted as pink, red, grey, purple, green, and yellow spheres, respectively.

Thus, the question of intraplanar magnetism is uniquely mapped to the question of structural arrangement. An empirical rule was suggested,[11] according to which all selenides always form the triangular structure (for undetermined reasons), while sulfides form a zigzag if the ionic radii of the two metals differ by less than 0.41 Å - 0.5 Å. This, however, is poorly supported by the evidence: using ionic radii from the WebElement's periodic table,[18] we find the differences shown in **Table 1**.

**Table 1. Ionic radii differences (Å) of several transition metals**

| | $V^{3+}$ | $Cr^{3+}$ | $Sc^{3+}$ | $In^{3+}$ |
|---|---|---|---|---|
| $Ag^{+}$ | 0.51 | 0.535 | 0.48 | 0.48 |
| $Cu^{+}$ | 0.13 | 0.155 | 0.10 | 0.10 |

As we see, $Sc^{3+}$ and $In^{3+}$, for instance, are barely distinguishable from $V^{3+}$ and $Cr^{3+}$, yet are structurally very different. Besides, it is completely unclear why the selenides behave differently. To address this question, we have performed DFT calculations of the energy difference between $AgVP_2S_6$ and $AgVP_2Se_6$. We found that the former has lower energy in the zigzag structure, whether the height of the chalcogen is relaxed or not. The corollary is that the structural energy difference is determined by the different electronic structure and not different size, *i.e.*, by the position of the chalcogen *p*-band.

Interlayer magnetic order is a completely different issue. Indeed, $AgVP_2Se_6$ is confirmed to be a ferromagnet in bulk with $T_C \approx 18$ K[7,8,15] while its Cr-based counterpart exhibits AFM interlayer coupling below 42 K.[11,19] Since the interlayer magnetic coupling in the quaternary TMCs is weak, it is sensitive to stacking registry: monoclinic $AgVP_2Se_6$ versus trigonal $AgCrP_2Se_6$, as is typical for van der Waals magnets (*e.g.* $CrI_3$,[20–22] where the FM interlayer magnetic order can be switched to AFM by relative shift on neighboring layers). The same trend appears in the CuM'$P_2S_6$ family: $CuVP_2S_6$ orders ferromagnetically[10] and $CuCrP_2S_6$ is an A-type antiferromagnet,[14] both with triangular V/Cr sublattice.[11]

The uniqueness of the FM state in $AgVP_2Se_6$ merits a deeper investigation of spin-phonon coupling in this compound. However, existing Raman studies have not explored the FM regime below the Curie temperature,[7,23] do not account for all predicted modes, and report conflicting symmetry assignments.[23] A comprehensive low-temperature Raman investigation capable of accessing the full phonon manifold, probing spin–phonon coupling in the FM regime, and resolving the outstanding mode assignments is thus called for. Here, we address this knowledge gap through a comprehensive temperature-dependent, magnetic field-dependent, and angle-resolved Raman spectroscopy (ARRS) of $AgVP_2Se_6$. We examine two types of crystals, CVT- and flux-grown, which results in different layer stacking[8] and thereby enables the study of interlayer coupling on phonon modes. We interpret measurements using DFT calculations of phonons, magnetism, and optical absorption.

Our measurements resolve up to 30 vibrational modes, enable phonon symmetry assignments, and reveal complex-valued Raman tensors characteristic of this low symmetry crystal. Spectral differences between CVT- and flux-grown crystals are consistent with the growth-influenced stacking disorder. Beyond establishing a comprehensive $AgVP_2Se_6$ phonon reference, our measurements reveal pronounced spectral changes near 100 K - well above the FM transition - pointing to a previously unreported structural phase transition that adds a new dimension to the phase diagram of this material. Another open question concerns the electronic structure of $AgVP_2Se_6$. Transport measurements yield an activation gap of ≈ 0.325 eV, yet optical absorption measurements place the band edge at ≈ 2.14 eV - a discrepancy far too large to be explained by the indirect gap argument typically invoked in such cases.[7] DFT band-structure calculations identify an isolated V-d band that is optically dark but thermally accessible, reconciling the long-standing activation gap discrepancy. Together, these results establish the foundational phonon properties of $AgVP_2Se_6$ and lay the groundwork for future studies of the potential of this material in spintronic applications and van der Waals heterostructures.

## RESULTS AND DISCUSSION

### Crystalline Structure and Stacking Domain Variants.

The bulk $AgVP_2Se_6$ crystal belongs to space group $C_2$ in a non-centrosymmetric monoclinic structure, with unit cell lattice constants of $\mathbf{a} = 6.3399$ Å, $\mathbf{b} = 11.020$ Å, $\mathbf{c} = 6.9819$ Å and $\beta = 106.824°$.[15] For the monoclinic $AgVP_2Se_6$ crystals, the unique ***b***-axis is perpendicular to ***a***- and ***c***- axes while ***a*** and ***c*** make an oblique angle $\beta$. Each layer which is built by the stacking of $PSe_3$ - V/Ag - $PSe_3$ units extends across the ***ab***-plane and stacks along the ***c***-axis by weak van der Waals interactions as shown in **Figure 2a**. Here, single crystals of $AgVP_2Se_6$ were synthesized via chemical vapor transport (CVT) and horizontal flux growth. We then mechanically exfoliated the bulk crystals using blue tapes and transferred the flakes onto $SiO_2$/Si substrates for measurements. A microscopic image of the exfoliated CVT-$AgVP_2Se_6$ flake is given in **Figure 2a**. In the previous work reported by Miao *et al.*, six different domain types along the ***z***-axis were predicted using group theory and experimentally observed using atomically resolved annular dark-field scanning transmission electron microscopy.[8] From the standard stacking domain, other variants are formed by an 120° in-plane rotation of adjacent layers and chirality switching of the Ag and V atoms.[8] Distinct types of domains boundaries can be formed when these domains interface with each other in the out of plane direction.[8] It is worth noting that a higher density of domain boundaries is observed in the flux-grown $AgVP_2Se_6$ crystal compared to the CVT-grown one.[8] Such structural complexity could impact properties associated with interlayer interactions and will be discussed later in this work. We first investigate the CVT-grown crystal that has comparatively fewer domain boundaries as determined by scanning transmission electron microscopy (STEM) images.[8]

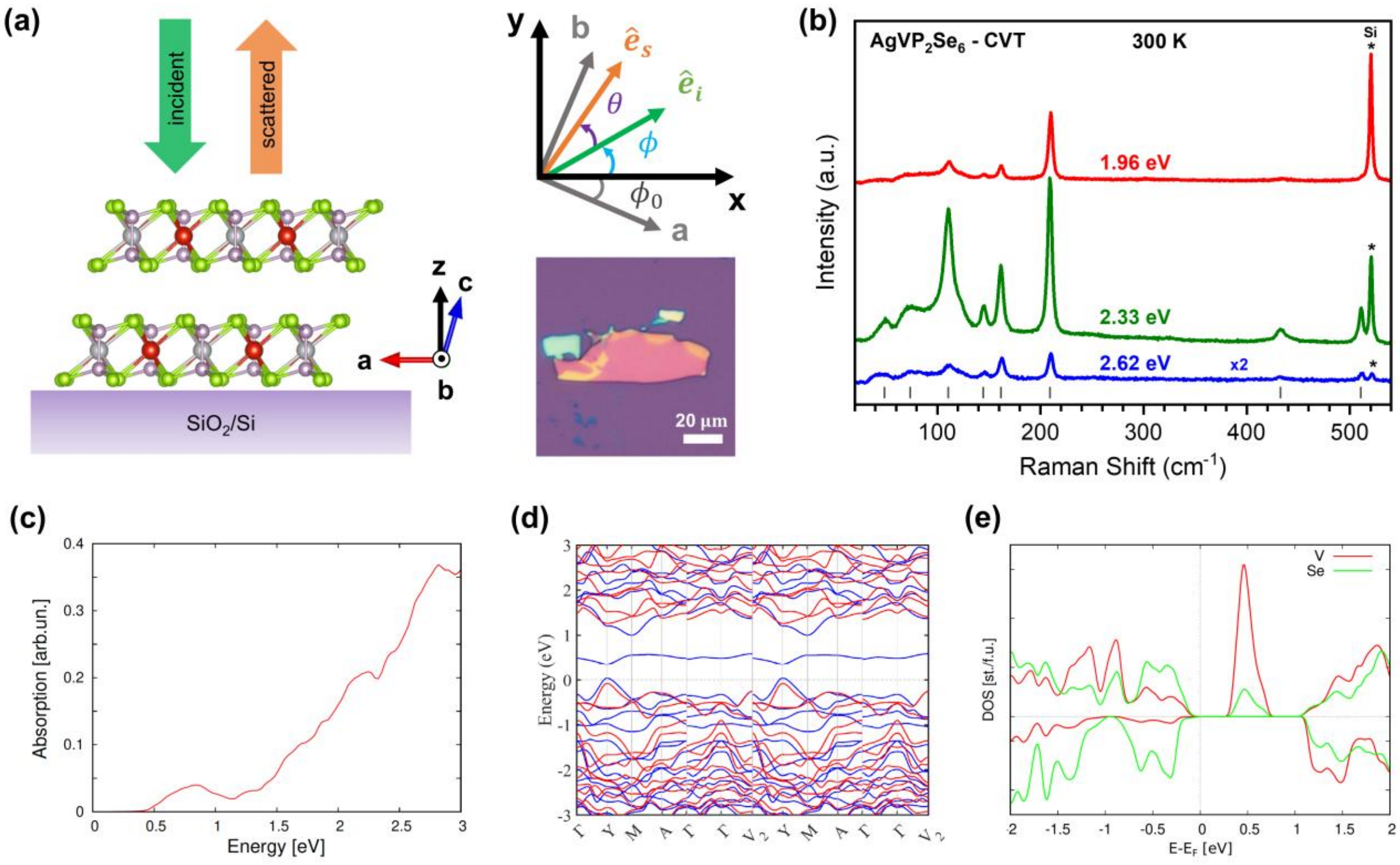

**Figure 2.** (a) Back-scattering measurement configuration of $AgVP_2Se_6$ flakes from the side view (left) and top view (right), along with a microscopic image of the CVT- $AgVP_2Se_6$ flakes. The crystalline ***ab***-plane is perpendicular to the incident light but noncollinear to the laboratory frame ***xy***-plane with an offset angle $\phi_0$. The initial orientation of incident polarized light is set along the ***x***-axis. The angle $\phi$ represents the sample rotation with respect to the ***x***-axis. $\theta$ is the angle between the incident ($\hat{\boldsymbol{e}}_i$) and scattered ($\hat{\boldsymbol{e}}_s$) light, which is fixed while $\phi$ varies. (b) Multi-excitation (1.96 eV, 2.33 eV, 2.62 eV) Raman spectra collected at room temperature from the exfoliated CVT-grown $AgVP_2Se_6$ flake. Spectra are collected at the same power and exposure time, showing a quasi-resonant effect at 2.33 eV. (c) DFT-calculated absorption curve for $AgVP_2Se_6$. (d) DFT-calculated band structure and (e) projected electronic density of states for $AgVP_2Se_6$. The mid-gap V band is responsible for the FM state and the difference in the activation energy and optical gap.

## Raman Resonance and Optical Gap

Our Raman characterizations are carried out in the back-scattering configuration illustrated in **Figure 2a.** Here, the crystalline ***ab***-plane is parallel with the ***xy***-plane in the laboratory coordination. The incident and scattered light propagate along the ***z-*** axis perpendicular to the cleavage plane. Their polarization orientations are represented by the unit vectors $\hat{e}_i$ and $\hat{e}_s$, respectively, and $\theta$ is the relative angle between them. This angle is fixed at $\theta = 0$ for co-polarized or $\theta = 90°$ for cross-polarized configuration. The initial orientation of the incident polarized light is set along the ***x***-axis. Since the exfoliated flakes are oriented in arbitrary directions, we denote an offset angle $\phi_0$ between the crystallographic ***a***, ***b*** axes and the lab frame ***x***, ***y*** axes, as shown in **Figure 2a**.

**Figure 2b** shows unpolarized room temperature Raman spectra from the exfoliated CVT-grown $AgVP_2Se_6$ flake on a $SiO_2/Si$ substrate, collected with 633 nm, 532 nm, and 473 nm laser wavelengths which correspond to energies of 1.96 eV, 2.33 eV, and 2.62 eV, respectively. Eight Raman modes are present in all three spectra, appearing at 49 $cm^{-1}$, 74.6 $cm^{-1}$, 110.7 $cm^{-1}$, 144.9 $cm^{-1}$, 161.8 $cm^{-1}$, 209.7 $cm^{-1}$, 433 $cm^{-1}$, and 511 $cm^{-1}$. The first seven peaks agree with prior room temperature Raman studies[7,23] while the additional peak at 511 $cm^{-1}$ arises from out-of-plane vibrations of P atoms and has also been detected in similar quaternary TMC compounds.[19] Intensities of all the Raman peaks depend strongly on the excitation energy, showing a quasi-resonant effect at 2.33 eV. This energy threshold well aligns with the optical transition ≈ 2.3 eV for $AgVP_2Se_6$ measured by Miao *et al.*[8] and therefore, we use 532 nm excitation for the entirety of this study.

The observation of a quasi-resonance behavior near the optical gap merits discussion as the value of 2.3 eV is quite different from the measured transport gap of 0.325 eV.[7] To address this confusion, we calculated the optical absorption spectrum of bulk $AgVP_2Se_6$ using DFT (**Figure 2c**) along with the full electronic band structure (**Figure 2d**). A well-defined gap of about 0.4 eV is clearly visible followed by weak absorption between 0.5 eV and 1.2 eV. Above 1.2 eV, the absorption rises quickly which is qualitatively consistent with experimental measurements in references,[7,8] except for the fact that the absorption increases at ≈ 2.14 eV. This discrepancy is expected as DFT calculations tend to overly delocalize orbitals which leads to a general underestimation of band gaps.

While the above results are consistent with experimental measurements of the optical gap, the transport gap of 0.325 eV remains unexplained. This apparent discrepancy is examined via DFT-calculated band structure in **Figure 2d**. Besides a direct gap of 1.3 eV at the Y points, there is also an additional single band sitting 0.35 eV above valence band maximum (VBM) which is the likely origin of the activation gap observed in transport. We present the elemental-resolved projected density-of-states in **Figure 2e** which shows that this isolated, relatively flat band is primarily of V-*d* character and likely responsible for the FM state. In contrast, the VBM is predominantly Se-*p* while the major conduction bands are a mixture of V-*d* and Se-*p*.

This band character profile provides a natural interpretation of the experimental findings. It explains why the optical absorption measurement is more sensitive to the larger gap between the major valence and conduction band, while the smaller activation gap, despite also being a direct gap, lies in the mid-infrared range beyond the detection range of common spectrometer systems. The optical absorption is directly related to the dipole matrix element $\langle c|\hat{P}_{dip}|v\rangle$ where $\hat{P}_{dip}$ is the dipole operator and $c$ and $v$ are the different eigenstates of the same momentum. The rapid increase in optical absorption at a photon energy of 1.3 eV corresponds to the larger gap because the major conduction band contains a significant proportion of Se-*p* character, which contributes strongly to the dipole matrix element. This might explain the strong energy excitation dependence of the major Raman peaks between 100 cm$^{-1}$ and 210 cm$^{-1}$ shown in **Figure 2b** as they correspond mainly to vibrations of the Se atoms. In contrast, the smaller gap, associated with the transition from Se-*p* to single band of predominantly V-*d* character, indicates that it is a charge-transfer insulating gap rather than a Mott insulating gap and the dipole matrix elements for transitions involving this band being relatively small. On the other hand, thermal excitations leading to electron transport behave differently. As a diffusion process, they are insensitive to the detailed character of the wavefunctions. The smaller gap defined by the single band therefore manifests in temperature-dependent resistivity measurements.

**Angle-Resolved Raman Spectroscopy (ARRS)**

The unit cell of $AgVP_2Se_6$ contains 10 atoms, resulting in 27 optical phonon modes that are all Raman active (13 A + 14 B) as predicted by DFT calculations. The symmetry of a given mode can be determined by measuring the Raman scattering intensity as a function of polarization angle. The Raman scattering intensity $I$ is given by

$$I \propto |\hat{e}_s . \boldsymbol{R} . \hat{e}_i|^2 \tag{1}$$

where $\boldsymbol{R}$ is the Raman tensor for a given mode, and $\hat{e}_i$ and $\hat{e}_s$, are the unit polarization vectors of the incident and scattered light, respectively. The general forms of Raman tensors for modes A and B of typical $C_2$-crystals with unique axis $\boldsymbol{b}$ are given in **Table S1**.[24–26] Readers should also be aware that these tensors differ if instead the convention with unique axis $\boldsymbol{c}$ is adopted.[24,27] Because of the strong optical absorption of $AgVP_2Se_6$, the Raman tensor elements must be treated as complex numbers,[25,28] and we use the following forms for A and B modes, respectively:

$$R_A = \begin{pmatrix} ae^{i\varphi_a} & 0 & de^{i\varphi_d} \\ 0 & be^{i\varphi_b} & 0 \\ de^{i\varphi_d} & 0 & ce^{i\varphi_c} \end{pmatrix} \tag{2}$$

$$R_B = \begin{pmatrix} 0 & fe^{i\varphi_f} & 0 \\ fe^{i\varphi_f} & 0 & ee^{i\varphi_e} \\ 0 & ee^{i\varphi_e} & 0 \end{pmatrix} \tag{3}$$

where $a$ - $f > 0$ and $\varphi_i$ ($i =$ $a$ - $f$) denote the phase of each tensor component. To experimentally determine mode symmetry and extract the Raman tensor elements defined above, we employed ARRS in the back-scattering geometry. ARRS also provides a reliable means of determining the in-plane crystal orientation when the crystallographic axes cannot be unambiguously inferred from flake morphology alone.[29,30] This is particularly relevant for $AgVP_2Se_6$, where stacking domain variants can further complicate the correspondence between visible flake cleaved edges and the underlying ***a-*** or ***b-*** axes.[8] In our setup, the sample remains stationary while the polarized incident laser is rotated by a half-wave plate (HWP) in a motorized mount placed after the edge filter and before the objective. The light polarization vector in the ***xy***-plane varies as a function of rotation angle $\phi$ cand can be expressed as $\hat{e}_i = [\cos(\phi + \phi_0), \sin(\phi + \phi_0), 0]$, with $\phi_0$ is the offset angle between $\hat{e}_i$ and the ***x***-axis when the HWP is at zero. Extracting the angle $\phi_0$ allows us to identify the crystal axis relative to the predetermined polarization direction of incident light which is set by a fixed polarizer. By using an analyzer in the detection path, we can determine the polarization of the Raman scattered photons. We set the analyzer to be either parallel (co-polarized, ∥) or perpendicular (cross-polarized, ⊥) to the incoming laser polarization. The polarization vector of the scattered light is therefore given by $\hat{e}_{s,\parallel} = [\cos(\phi + \phi_0), \sin(\phi + \phi_0), 0]$ or $\hat{e}_{s,\perp} = [-\sin(\phi + \phi_0), \cos(\phi + \phi_0), 0]$ for parallel and perpendicular cases, respectively. Substituting eqs 2 and 3 into eq 1 yields the expected angle-dependent Raman intensity for each mode and polarization configuration:

$$I_A^{\parallel} \propto a^2 \cos^4(\phi + \phi_0) + b^2 \sin^4(\phi + \phi_0) + 2ab \sin^2(\phi + \phi_0) \cos^2(\phi + \phi_0) \cos \varphi_{ab} \quad (4)$$

$$I_A^{\perp} \propto \sin^2(\phi + \phi_0) \cos^2(\phi + \phi_0) \, (a^2 + b^2 - 2ab \cos \varphi_{ab}) \quad (5)$$

$$I_B^{\parallel} \propto f^2 \sin^2 2(\phi + \phi_0) \quad (6)$$

$$I_B^{\perp} \propto f^2 \cos^2 2(\phi + \phi_0) \quad (7)$$

where $\phi_{ab} = \phi_a - \phi_b$ is the relative phase between the tensor components, which vanishes for a real-valued tensor and cancels entirely for B modes.

The above equations show that A and B modes appear in both configurations but with different periodicities and intensities that vary substantially with laser polarization and analyzer orientation. Therefore, we average the parallel and perpendicular data over all angular orientations to maximize signal-to-noise and ensure the identification of all active Raman modes (**Figure 3a**).[31] Here, we fit each peak using Lorentzian function and the Raman intensity refers to the fitted area throughout the paper. While only 8 modes were observed at room temperature (**Figure 2b**), cryogenic Raman measurements significantly enhanced the signal strength and reduced the linewidth resulting in the observation of 30 peaks in the range of 30 $cm^{-1}$ to 560 $cm^{-1}$, far more than prior studies.[7,23] Polarization-resolved spectra further separate several closely spaced modes near 146 $cm^{-1}$, 163 $cm^{-1}$, 435 $cm^{-1}$ and 444 $cm^{-1}$ (**Figure S1**). The number of observed peaks exceeds the 27 modes predicted by DFT, suggesting the presence of higher-order Raman modes from multi-phonon scattering. Weak and broad features around 463 $cm^{-1}$, 478 $cm^{-1}$ and 490 $cm^{-1}$ resemble second-order overtones of the fundamental modes at 231.6 $cm^{-1}$, 237 $cm^{-1}$ and 244 $cm^{-1}$, respectively, though combination Raman modes cannot be ruled out. Another possibility is the emerging vibrational modes activated by stacking disorder, which we will discuss later when comparing CVT and flux grown samples. **Table S2** compares the calculated and experimentally observed mode frequencies. The overall agreement is reasonable, though DFT-calculated frequencies (denoted as hollow triangles at the bottom of **Figure 3a**) are consistently offset from the measured values, indicating a uniform scaling discrepancy. Scaling the DFT results by 5%, as presented in **Figure 3a**, largely closes this gap for modes below 250 $cm^{-1}$, while higher-frequencies modes associated with P atoms vibration still show considerable mismatch. This discrepancy likely reflects DFT's tendency to overly soften phonon modes due to screening effects, resulting in systematically underestimated frequencies.[32,33] **Table S3** displays calculated phonon eigenvectors describing the atomic displacements for the Raman modes. The five modes above 400 $cm^{-1}$ are purely associated with P atoms, with the highest one corresponding to the out-of-plane vibration and the remaining four being in-plane. Interestingly, a mode predicted at 314.8 $cm^{-1}$ and observed at 320 $cm^{-1}$ shows strong hybridization between P and V, and is absent in the sister compound $AgCrP_2Se_6$.[19] Three modes in the 200 $cm^{-1}$ range are dominated by $V^{3+}$ ions vibration. Most modes below 70 $cm^{-1}$ primarily involved Se motion, except for the 101 $cm^{-1}$ mode whose eigenvector shows significant $Ag^+$ displacement. The three lowest-frequency modes correspond to Ag in-plane and out-of-plane vibrations, distinguishing $AgVP_2Se_6$ from $AgCrP_2Se_6$ where the $Ag^+$ ions instead exhibit a helical vibrational mode.[19]

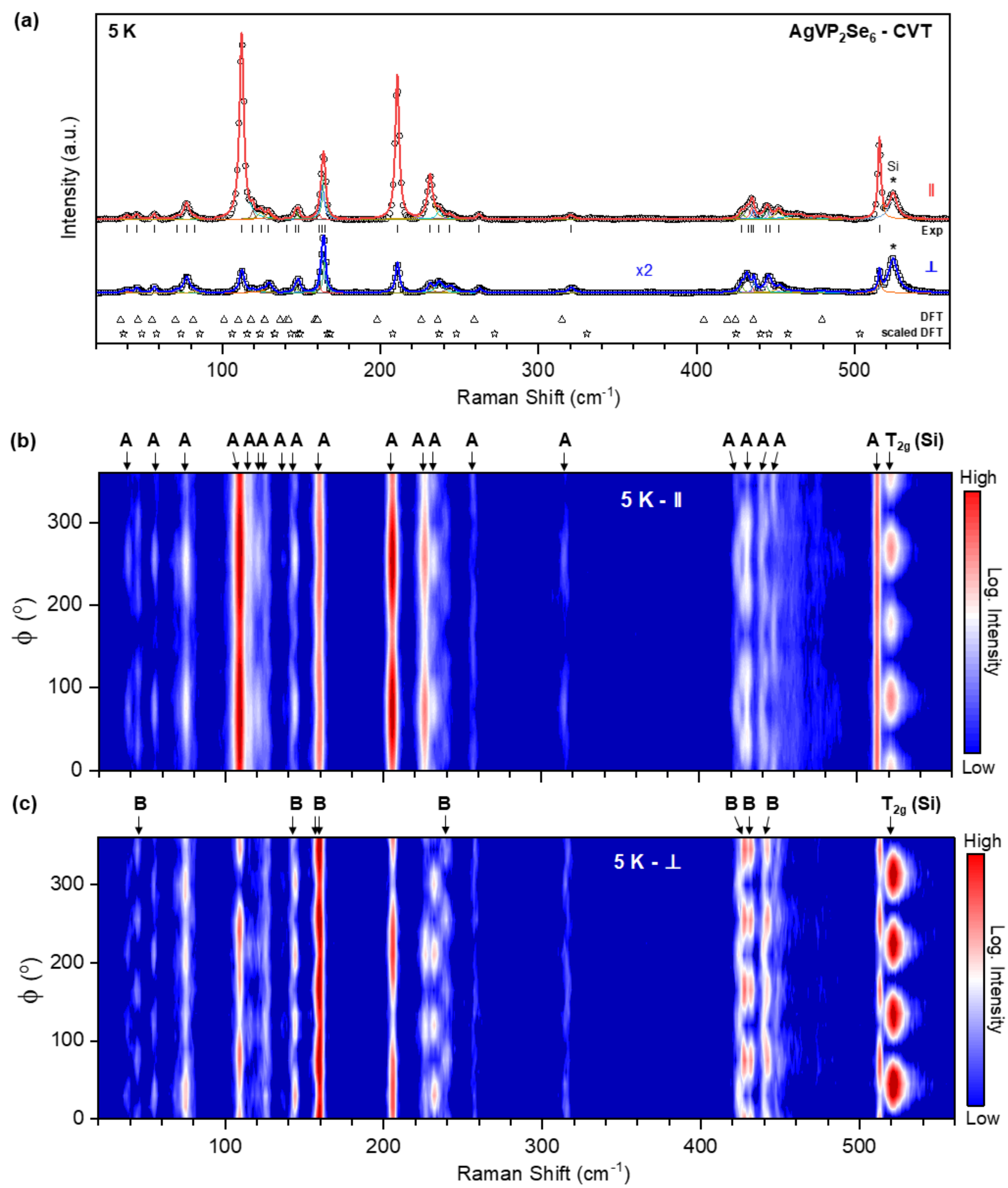

**Figure 3.** (a) Polarized Raman spectra of the CVT-grown $AgVP_2Se_6$ flake averaged over all crystal orientations for co-parallel (‖, circles) and cross-parallel (⊥, squares) configurations at 5 K using 532 nm laser. The solid lines represent a multi-Lorentzian fit, capturing the presence of all 30 Raman modes. Hollow triangles (bottom) represent 27 DFT-calculated Raman active modes, while the stars represent predicted frequencies scaled by 5%. ARRS maps of $AgVP_2Se_6$ illustrated by false-color graphs on a logarithmic intensity scale are presented in the (b) ‖ and (c) ⊥ polarization configurations.

While the frequency comparison between experiment and DFT is largely favorable, assigning Raman mode symmetries requires examining ARRS maps presented in **Figs. 3b** and **3c** for the parallel and perpendicular configurations, respectively. We assemble these maps by setting the polarizer/analyzer orientation and rotating the half wave plate (HWP), which is equivalent to physically rotating the sample itself. Subsequently, we can determine the symmetry of each mode by examining plots of Raman intensity versus angle and fitting this data to eqs 4 – 7. The angle dependent intensity equations determined from the $C_2$ Raman tensors indicate that A modes will exhibit two (four)-fold symmetry for ∥ (⊥) measurements while B modes exhibit four-fold symmetry pattern in both polarization configurations.

Representative polar plots of the angle-dependent Raman intensities $I_{\parallel}$ and $I_{\perp}$ for selecting A- and B-symmetry modes are shown in **Figure 4**, with the remaining modes presented in **Figures S2 and S3**. The solid lines are fitted to eqs 4 -7, which reproduce the angular dependence of all measured modes. To determine the crystal orientation from the ARRS measurement, we first look at the maximum intensity of A modes in the parallel polarization, as indicated by the red dash lines pointing to $\phi = 75°$ and $\phi = 255°$ in **Figure S4**. The fitting parameters include the initial offset angle value $\phi_0$ for each mode, with an average value around 15°. Therefore, one could easily estimate that the maximum of intensity occurs when the incident light polarization is at $\phi + \phi_0 = 90°$ or $270°$ or perpendicular to the crystal ***a***-axis, as reported in other monoclinic systems.[27,34] The near identical orientation of polar plots for A and B modes suggests that the crystallographic ***a***-axis makes an angle of $\phi_0 \approx 15°$ with the lab frame ***x***-axes or the initial direction of the polarized incident light, and we can incorporate $\phi_0$ as a constant into all fitting functions for simplicity.

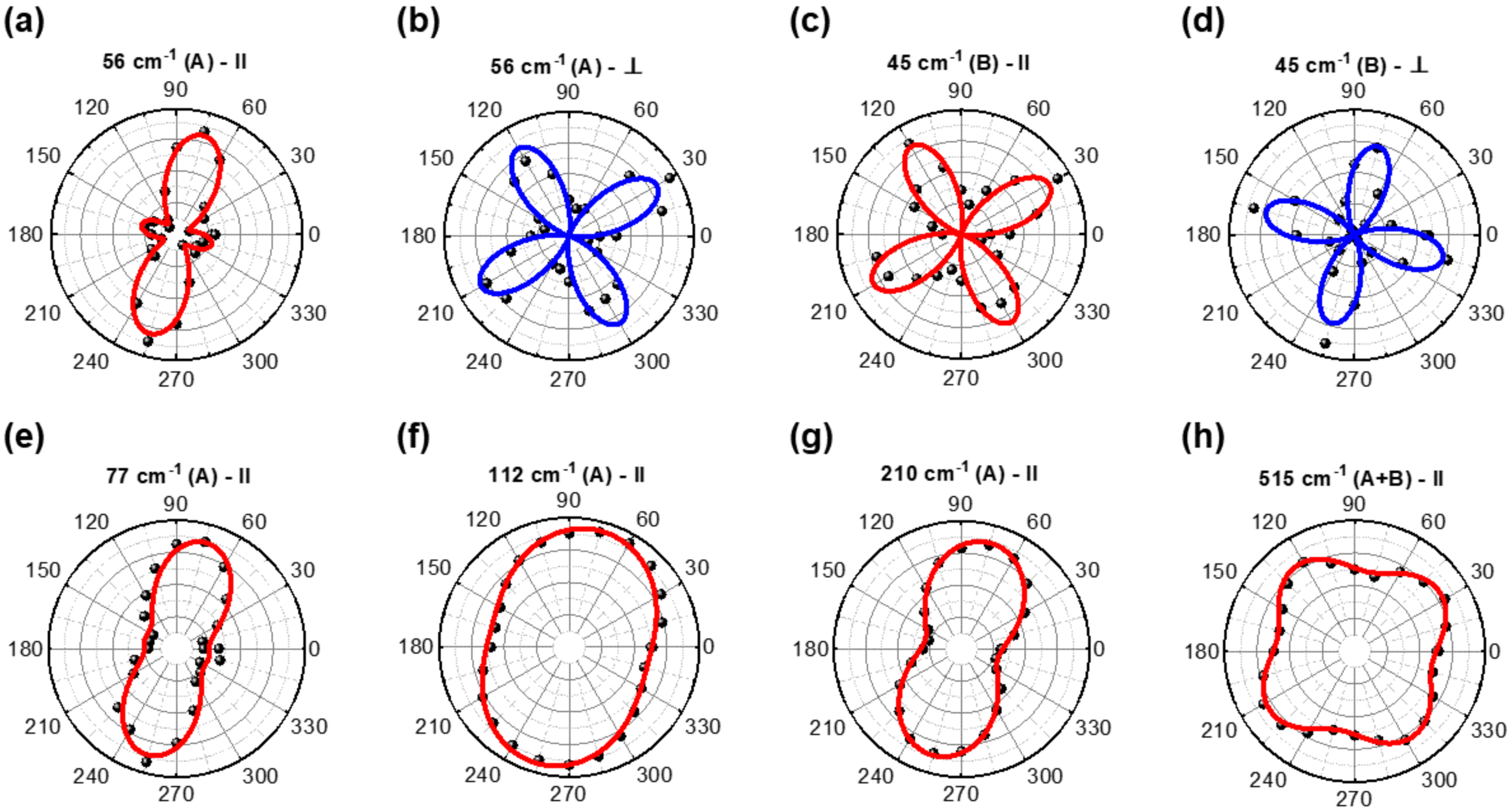

**Figure 4.** Polar plots of the $\phi$ angle-dependent Raman fitted area in the co- (∥,red) and cross- (⊥, blue) polarization configurations for identifying A- and B-modes at 5 K. Black spheres are measured Raman fitted area and solid lines are fits using eqs 4-7 as appropriate. A combined function, $I_A^{\parallel} + I_B^{\parallel}$, is required to fit the polar plot of the 515 cm$^{-1}$ mode. See **Figures S2** and **S3** for similar fits to other Raman modes which form the basis of our symmetry assignments.

Understanding the distinctive angle dependence of the Raman intensity of $AgVP_2Se_6$ requires complex tensor analysis. Although all A modes show 180° (or 90°)- periodicity in co (or cross)- parallel configuration (**Figures 4a, b)**, the shape of their polar plots is dictated by the tensor element *a, b* and the complex phase $\varphi_{ab}$ (**Figures 4e-h**). The 56 cm$^{-1}$ mode illustrates this clearly: its angular dependence in the parallel configuration, $I_A^{\parallel}(\phi)$, can only be reproduced by including the complex tensor elements $\varphi_{ab} \neq 0$ (**Figure S4b**), while a real-valued tensor fit fails entirely (**Figure S4a**). The key signature of the complex phase is the emergence of secondary maxima at 165° and 345°, a feature also reported in other low-symmetry orthorhombic and monoclinic crystals.[25,27,34] By fitting the data in **Figure S2** to eqs 4 and 5**,** we can extract the *a/b* ratio and relative phase $\varphi_{ab}$ for all the A modes as summarized in **Table S4**. Notably, the relative phase $\varphi_{ab}$ deviates from zero for the majority of A modes (**Table S4**), confirming the complex nature of the Raman tensor in this low-symmetry crystal.[25,27] In contrast, the symmetric four-lobe polar plots of all B modes illustrate the lack of a complex phase factor in eqs 6 and 7 (**Figures 4c, d, and S3**). For the 146 $cm^{-1}$, 163 $cm^{-1}$, 436 $cm^{-1}$, and 446 $cm^{-1}$ modes, a pure B-mode equation could not adequately fit the data; instead, a combination $I_A + I_B$ was required, suggesting possible overlap with neighboring modes of A-symmetry. This mixing produces the characteristic opening of the four-lobed polar plots for these modes in both ∥ and ⊥ configurations (**Figure S3**).

A similar approach can be applied to explain the anomalous behavior of the A-mode peak at 515 $cm^{-1}$ of $AgVP_2Se_6$ and the Si substrate $T_{2g}$-mode peak at 524 $cm^{-1}$. The former one is distinctive for its 90°-periodicity in co-parallel configuration (**Figures 4f, S5a**). Interestingly, the angle-dependent Raman intensity of the 515 $cm^{-1}$ mode can be fitted using $I_A^{\parallel} + I_B^{\parallel}$ and $I_B^{\perp}$ in the co- and cross-polarization, respectively. In the case of Si (100) substrate, the angle-resolved Raman intensity calculated for the $T_{2g}$ mode (see **Supporting Information**, eqs 2,3) results in a similar function as of B-mode in eqs 6 and 7.[35] The extracted fitting parameters given in **Table S5.** The near-isotropic response of the 515 $cm^{-1}$ mode in $AgVP_2Se_6$ is associated with near the unity ratio *a/b* and $\varphi_{ab} = 0$ in the $I_A^{\parallel}$ term and results in the suppression of the A-mode contribution in the cross configuration. Its offset angles ($\phi_0$) are much smaller than the ones for Si substrate ($\phi_0^{Si}$), ruling out the possibility of the Raman leakage between these neighboring modes, but rather suggesting the overlapping of two intrinsic A- and B-modes of $AgVP_2Se_6$. On the other hand, the four-fold symmetry of Si $T_{2g}$-mode collapses to two-fold in the parallel configuration, which we attempt to fit with $I_{T_{2g}} + I_A$ functions (**Figure S5a**) and fully recover in the cross configuration (**Figures S5d**). Although the Si substrate's intrinsic $T_{2g}$ tensor is unaffected by the flake - the crystallographic offset extracted from bare and flake-covered regions of the same wafer is unchanged ($\phi_0^{Si} \approx 49°$, see **Supporting Information**, eqs 1-9 and **Table S5**) - we attribute this asymmetric, parallel-only polarization response to $AgVP_2Se_6$ acting as a polarization-dependent optical filter on light reaching and returning from the substrate. $AgVP_2Se_6$ exhibits strong in-plane optical anisotropy, evidenced by its layer-independent anisotropic second-harmonic generation,[23] consistent with a complex refractive index that differs along the ***a***- and ***b***-axes. This anisotropy is expected to modulate the substrate signal through the interference mechanism established for isotropic 2D overlayers on $SiO_2$/Si,[36] now rendered polarization-dependent, consistent with recent multilayer transfer-matrix treatments of ARRS for anisotropic flakes.[37] Because this filtering samples the same axis on both incident and collected paths in parallel configuration but orthogonal axes in cross configuration, the distortion appears only in parallel measurements (see **Supporting Information**, eqs 1-9 and **Table S5** for full derivation and supporting fit parameters).

In total, 20 A and 8 B modes were identified among the phonon modes of CVT-grown $AgVP_2Se_6$. Several of our symmetry assignments - for the modes at 77 $cm^{-1}$, 112 $cm^{-1}$, and 210 $cm^{-1}$ - contradict a recent ARRS report on CVT-grown $AgVP_2Se_6$ which classified these as B-type.[23] Some predicted B-type modes were not observed, likely due to their low intensity, which is consistent with its Raman response being governed by off-diagonal tensor elements and polarization selection rules.[25,27]

## Temperature- Dependent and Magneto-Raman Spectroscopy

Having established the low temperature phonon characteristics of CVT-grown $AgVP_2Se_6$, we now examine the impact of temperature and magnetic field on lattice dynamics. Though Chen *et al.* measured temperature-dependent Raman of a 2D $AgVP_2Se_6$ sheet, their results are limited to the high temperature regime above 200 K.[23] Our cryogenic capability, on the other hand, allows us to investigate the evolution of the Raman signals from $AgVP_2Se_6$ flakes across the Curie temperature ($T_C \approx 18$ K). **Figure 5a** shows Raman spectra obtained by heating an exfoliated flake from 1.8 K to 290 K in a ∥ polarization configuration. The decrease of peak intensity, increase of peak width, and the red shift of major modes can be conceptually accounted for by a combination of volumetric expansion and optical phonon decay.[23,38] We find the remarkable result that modes at 231 $cm^{-1}$ , 237 $cm^{-1}$, 244 $cm^{-1}$ and 262 $cm^{-1}$ vanish at temperatures above 100 K. In **Figure 5b,** we examine this effect by extracting the Raman integrated area from corresponding spectral regions (labeled as i – vi) in **Figure 5a**, normalizing them by the 1.8 K data, and plotting them as a function of temperature. Interestingly, integrated spectral area of modes in regions ii and iv, which are associated with Ag and V atoms, show a drastic drop centered around 100 K and reach a reduction of 70 % and 90 %, respectively, at 150 K as shown in **Figure 5b**. On the contrary, the change is insignificant in other spectral regions. A recent study of $AgCrP_2S_6$ shows a similar drop in phonon-mode count around 140 K to 200 K.[12] The disappearance of these modes is attributed to the symmetry-lowering structural transition from $C_{2h}$ point group at high temperatures to $C_2$ or $C_s$ point group at low temperatures.[12] Additionally, ARRS measurements at 5 K, 30 K, and 150 K show the polarization response of the 515 $cm^{-1}$ peak evolves from quasi-isotropic ($a \approx b$) to its A-type nature with two-fold symmetry ($a < b$) with increasing temperature (**Figure 5c** and **Table S5)**. Similar behavior is also shown in several modes of $AgCrP_2S_6$.[12] On the other hand, the anisotropic polarization pattern of Si in the ∥ configuration persists upon heating up to 150 K as shown in **Figures S5 a-c**. Intriguingly, at all temperatures, both 515 $cm^{-1}$ and Si peaks maintain their perfect four-fold symmetry in the ⊥ polarization (**Figures S5 d-f)**. In another study of the selenium analogue $AgCrP_2Se_6$, Susner *et al.* report a secondary phase transition near 120 K using temperature-dependent XRD and Raman spectroscopy which relates to the thermally induced distortion of the lattice structure.[11] All together, these anomalous temperature dependences of the phonon modes suggest that the structural instability may be a common feature of quaternary TMC compounds. Particularly in the case of $AgVP_2Se_6$, we also consider a distinct possibility: a temperature-driven transition caused by different stacking-domain variants[8], analogous to the monoclinic – orthorhombic structure transition in layered T'-$MoTe_2$.[39] This is plausible here because CVT- and flux-grown flakes, which host different stacking domains, show measurably different Raman spectra at low temperature as we will discuss later, indicating that domain population genuinely modulates the phonon spectrum of $AgVP_2Se_6$. However, this scenario calls for more thorough theoretical simulations and temperature-dependent structural characterizations beyond the scope of this work.

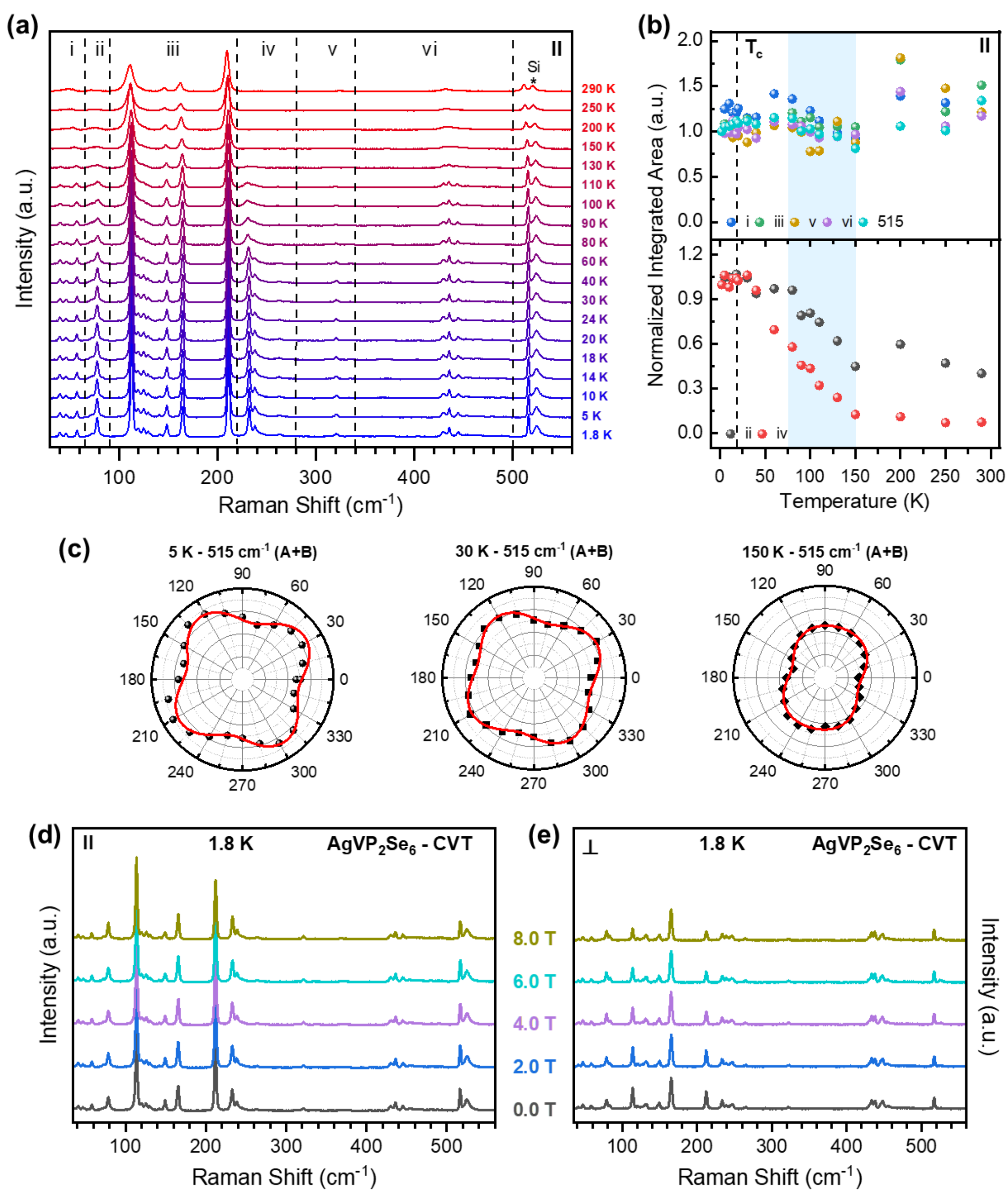

**Figure 5.** (a) Temperature-dependent Raman spectra collected on the CVT-grown $AgVP_2Se_6$ flake from 1.8 K to 290 K, in co-parallel polarization. The spectra are divided into six sections highlighted by dashed lines. (b) Integrated spectral area of region i-vi, along with fitted area 515 $cm^{-1}$ peak as a function of temperature are normalized to the Si peak then divided by the values at 1.8 K. (c) ARRS fitted area of the 515 $cm^{-1}$ mode in ∥ configuration, measured at 5 K, 30 K, and 150 K. (d) Co-polarized and (e) cross-polarized magneto-Raman spectra collected at 1.8 K at selected magnetic field from 0 T to 8 T, applied parallel to the direction of light propagation. Measurements were performed up to 9 T.

Temperature- and magnetic-field-dependent Raman spectroscopy are highly effective at identifying spin excitations (i.e. magnons), spin-phonon coupling, and magnon-phonon hybridization in TMCs such as $FePS_3$,[3] $MnPSe_3$,[2,40] and $FePSe_3$.[4,41] However, we find no significant discontinuities in the linearly polarized Raman spectra across the ferromagnetic transition at 18 K (**Figures 5a**) which was also confirmed by temperature-dependent ARRS maps measured below and above $T_C$ (**Figures S7a, b**). This implies a weak or even absent spin-phonon coupling in this material, which is further confirmed by magneto-Raman measurements. As the external magnetic field in the Faraday configuration is swept from 0 T to 9 T, linearly polarized Raman spectra of $AgVP_2Se_6$ remain nearly unchanged (representative spectra in **Figures 5d, e**), suggesting no magnons or hybridized magnon-phonon modes.

### Raman Assessment of Crystal Growth Methods

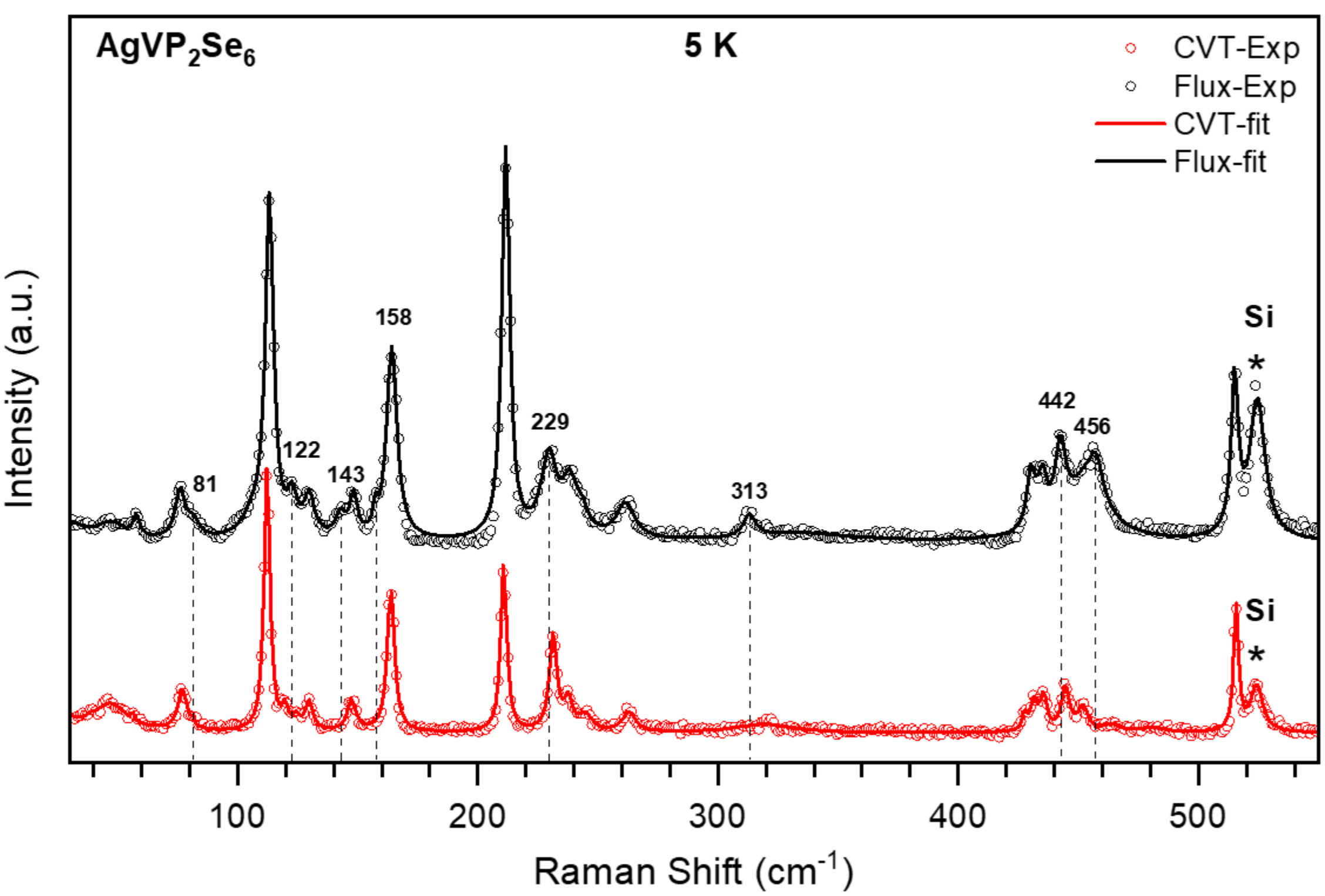

**Figure 6.** Raman spectra of CVT- and flux-grown $AgVP_2Se_6$ flakes at 5 K collected without polarization optics using 532 nm excitation. Dash lines highlight the emerging or shifted peaks in flux- (black) and CVT- (red) spectra. The numbers indicate Raman frequencies at the marked positions.

Raman signals are sensitive to the crystal structure and can be used effectively to evaluate the quality of the crystal growth process. **Figure 6** shows the comparison between the well characterized phonon Raman modes from the CVT- $AgVP_2Se_6$ and the spectra obtained from exfoliated flux-grown $AgVP_2Se_6$ flake. Both crystals can host different domain variants as described in prior work by members of our team, however, crystals grown by the flux method exhibit more stacking domains of different types as compared to the CVT crystals as demonstrated by high-resolution STEM images in the Miao *et al.* study.[8]

A noticeable feature is that the contrast difference between the spectra of these samples is presented in the high energy regime between 220 $cm^{-1}$ and 460 $cm^{-1}$. For example, the 231 $cm^{-1}$ peak appears sharper and its intensity dominant over the higher energy tail peaks (237 $cm^{-1}$ and 244 $cm^{-1}$) in the CVT-grown flake. The 320 $cm^{-1}$ -mode in CVT-spectra experiences a large shift to 313 $cm^{-1}$ in the flux-spectra, while a new peak at 455 $cm^{-1}$ prominently appears in the latter (highlighted by dash lines in **Figure 6**). The broadening of 77 $cm^{-1}$-mode in the Raman signal of flux-grown $AgVP_2Se_6$ shows another peak emerging on the higher energy side which was hardly detectable in the CVT-counterpart. Interestingly, most of these modified modes also disappear with the rising temperatures above 100 K, as discussed earlier. These overlapping observations strengthen our hypothesis of a new structural phase transition influenced by the complex domain and stacking variation in $AgVP_2Se_6$ synthesized by both CVT and flux-growth technique. More broadly, these results demonstrate that growth conditions can substantially modify the out-of-plane domain structure, thereby strongly impacting the Raman response and lattice vibrational properties.

## CONCLUSIONS

This work provides a low-temperature comprehensive spectroscopic and theoretical description of $AgVP_2Se_6$, a non-centrosymmetric monoclinic (C2) layered TMC in which all 27 zone-center optical phonons are Raman active. Cryogenic Raman measurements substantially enhance spectral signals, revealing up to 30 peaks (including candidate higher-order features) and enabling symmetry-resolved assignments beyond what is accessible at room temperature. ARRS measurements can only be captured with complex-valued Raman tensors, underscoring the importance of absorption effects in this low-symmetry crystal. The angular analysis also provides a practical route for crystal orientation determination and for interpreting anomalous polarization patterns arising from neighboring-mode mixing and substrate interference effects. Temperature-dependent Raman and magneto-Raman spectra suggest weak or absent spin–phonon coupling. We observed evidence for a previously unobserved structural phase transition near 100 K in the form of mode vanishing and polar plot changes. Comparisons between CVT- and flux-grown flakes indicate growth-dependent and visible changes consistent with enhanced stacking disorder in flux samples. Finally, DFT band-structure and orbital-character analysis reconciles the gap discrepancy by distinguishing the dominant direct optical transition (≈ 1.3 eV) from a much smaller activation gap (≈ 0.325 eV) associated with a separate, predominantly V–*d* derived band with weaker optical matrix elements. These results resolve key discrepancies surrounding the vibrational modes of $AgVP_2Se_6$ and reveal new questions surrounding its phase diagram.

## METHODS

### Growth of $AgVP_2Se_6$ single crystals and sample preparation

Chemical vapor transport (CVT) growth of $AgVP_2Se_6$ single crystal follows the procedure reported in the previous study.[7] However, the process is hindered by the formation of P–Se glass and $VSe_2$ secondary phases, yielding small crystals. To overcome this, single crystals of $AgVP_2Se_6$ were synthesized using horizontal flux method, as described in the previous work.[8] V, Ag, P, and Se powders were mixed homogeneously in a molar ratio of 1:1:2:6 and pressed into a 1-inch-diameter pellet, which was used as the precursor for $AgVP_2Se_6$ single crystal growth. The pellet was then cracked, and a portion was sealed in a quartz tube together with 5 g of the $AlCl_3$–KCl eutectic mixture as the flux in a molar ratio of 2:1. The vacuum-sealed ampoule was placed in a two-zone horizontal furnace, with the precursor positioned at the hot end maintained at 400 °C and the cold end maintained at 375 °C. After 8 days of growth, the furnace was naturally cooled to room temperature. Dark silver-colored crystals were observed in the cold zone. The crystals were washed first with deionized water and then with acetone, yielding plate-like $AgVP_2Se_6$ single crystals with submillimeter lateral dimensions. Bulk $AgVP_2Se_6$ single crystals, both CVT- and flux-grown, were mechanically exfoliated and transferred on $O_2$ plasma cleaned $Si/SiO_2$ substrates using blue tape.

**Raman Spectroscopy**

Angle-resolved Raman spectroscopy (ARRS) measurements of the $AgVP_2Se_6$ flakes were carried out in the first setup using a closed-cycle He-cooled cryostat with temperatures from 5 K to 300 K. The samples are excited in a backscattering geometry with 532 nm laser excitation focused through a 0.6 NA objective lens with 40x magnification. The scattered light was collected through the same lens and directed to a spectrometer with liquid-nitrogen cooled charge-coupled device detector and a diffraction grating with 1200 lines/mm. The rejection of the Rayleigh scattered light is accomplished using three-volume Bragg grating notch filters that enable the acquisition of data down to 10 $cm^{-1}$. The polarization optics include two linear polarizers (P1, P2) and a half-wave plate (HWP) in a motorized mount placed after the edge filter and before the objective. The co- and cross-parallel polarization configurations are achieved by rotating the polarizer P2 in the collection path while fixing the polarizer P1 in the excitation path. The laser power was measured to be 300 μW right before the objective.

The multi-excitation Raman spectroscopy was obtained at room-temperature in the second setup using a customized LabRAM HR Evolution confocal Raman microscope (HORIBA France SAS). Excitation involves three different laser wavelengths ($\lambda$ = 473 nm, 532 nm, 633 nm) focused through an objective, with the scattered light collected in a back-scatter geometry. The objective lens has 0.75 NA and 100x magnifications. The spectrometer's grating is 1,800 lines per mm.

Temperature- and magnetic-field-dependent Raman measurements were performed in the third setup using 532 nm excitation and a triple grating spectrometer (Horiba JY T64000, 1800 groove/mm grating), which is coupled to a liquid-nitrogen cooled CCD detector. Here the temperature can vary from 1.8 K to 290 K. The sample was placed into an attoDRY 2100 cryostat (Attocube Inc.). The magnetic field is applied perpendicular to the sample surface, in fields up to 9 T. The excitation laser is focused on micrometer-sized flakes with a low-temperature, magnetic field compatible microscope objective (50x, N.A. = 0.82), with laser spot sizes near 1 $\mu$m. The magneto-Raman measurements acquired were automatically corrected for Faraday rotation in the objective using two motorized achromatic half-wave plates (HWP1, HWP2), and two fixed polarizers (P1, P2) external to the magneto cryogenic system. The laser power was set to 100 μW -300 μW using ND filters to avoid overheating or damage on the surface of the flakes.

**DFT Calculations**

Electronic structure calculations were performed using Vienna ab initio Simulation Package (VASP)[42] within projector augmented wave (PAW) method.[43] The Perdew-Burke-Ernzerhof (PBE)[44] generalized gradient approximation was employed to describe exchange correlation effects. For the ordered states, we added a Hubbard U correction with the fully localized limit double-counting recipe,[45,46] to account for the strongly correlated V-3d states and their localized magnetic moments. The effective parameter U-J=2 eV was used. The zone-center phonon modes involved in Raman spectra were calculated using DFPT routine (IBRION=8) implemented in VASP.

## ASSOCIATED CONTENT

**Supporting Information**

General Raman tensor forms for the C2 point group and derivations underlying the complex-tensor analysis; measured versus calculated Raman peak frequencies and symmetry assignments; DFT-calculated atomic displacements; fitted tensor parameters of A modes retrieved from co-polarized data; additional polarization-resolved Raman spectra; angle-dependence Raman intensity plots of A- and B- modes; and angle-resolved Raman polar maps at different temperature and the anomalous substrate polar plots.

## AUTHOR INFORMATION

### Corresponding Authors

**Patrick M. Vora** – Department of Physics and Astronomy, George Mason University, Fairfax, VA 22030, United States; orcid.org/0000-0003-3967-8137; Email: pvora@gmu.edu

**Thi Hai Yen Pham** – Department of Physics and Astronomy, George Mason University, Fairfax, VA 22030, United States; orcid.org/0000-0003-1755-4117; Email: tpham62@gmu.edu

### Authors

**Po-hao Chang** – Department of Physics and Astronomy, George Mason University, Fairfax, VA 22030, United States; orcid.org/0000-0003-0444-4672

**Aidan C. Malloy** – Department of Physics and Astronomy, George Mason University, Fairfax, VA 22030, United States

**Seng Huat Lee** – 2D Crystal Consortium, Materials Research Institute, The Pennsylvania State University, University Park, PA 16802, United States; orcid.org/0000-0003-4254-3460

**Zhiqiang Mao** – 2D Crystal Consortium, Materials Research Institute, The Pennsylvania State University, University Park, PA 16802, United States; orcid.org/0000-0002-4920-3293

**Maria F. Munoz** – Quantum Measurement Division, Physical Measurement Laboratory, NIST, Gaithersburg, MD 20899, United States; orcid.org/0000-0003-1525-2318

**Riccardo Torsi** – Quantum Measurement Division, Physical Measurement Laboratory, NIST, Gaithersburg, MD 20899, United States; orcid.org/0000-0001-7748-1074

**Angela R. Hight Walker** – Quantum Measurement Division, Physical Measurement Laboratory, NIST, Gaithersburg, MD 20899, United States; orcid.org/0000-0003-1385-0672

**Igor Mazin** – Department of Physics and Astronomy, George Mason University, Fairfax, VA 22030, United States; orcid.org/0000-0001-9456-7099

### Notes

The authors declare no competing financial interests.

## ACKNOWLEDGEMENTS

The materials synthesis is conducted at The Pennsylvania State University Two-Dimensional Crystal Consortium-Materials Innovation Platform (2DCC-MIP), which is supported by the U.S. National Science Foundation (NSF) cooperative agreement DMR-1539916. P.M.V and T.H.Y.P acknowledge support provided by NSF through grant no. 2226097. P.C and I.M were supported by the NSF under award no. DMR-2403804. The authors would like to thank T. Adel for assistance with Raman measurements at NIST.

### NIST disclaimer

Commercial equipment, instruments, and materials are identified in this paper to adequately specify the experimental procedure. Such identification is not intended to imply recommendation or endorsement by the National Institute of Standards and Technology or the United States Government, nor is it intended to imply that the materials or equipment identified are necessarily the best available for the purpose.

This document has not been peer reviewed but has been cleared by NIST for release.

**Supporting Information for**

# Unveiling Lattice Dynamics and a Hidden Structural Transition in the 2D Ferromagnet $AgVP_2Se_6$ via Raman Spectroscopy

Thi Hai Yen Pham[1,2*], Po-hao Chang[1,2], Aidan C. Malloy[1,2], Seng Huat Lee[3], Zhiqiang Mao[3], Maria F. Munoz[4], Riccardo Torsi[4], Angela R. Hight Walker[4], Igor Mazin[1,2], Patrick M. Vora[1,2*]

[1]Department of Physics and Astronomy, George Mason University, Fairfax, VA 22030

[2]Quantum Science and Engineering Center, George Mason University, Fairfax, VA 22030

[3]2D Crystal Consortium, Materials Research Institute, The Pennsylvania State University, University Park, PA 16802

[4]Quantum Measurement Division, Physical Measurement Laboratory, NIST, Gaithersburg, MD 20899

[*]Correspondence: tpham62@gmu.edu and pvora@gmu.edu

**Table S1. General Raman tensors of monoclinic crystals in $C_2$ point group**. The tensor components are complex numbers.

| Mode | A | B | |
|---|---|---|---|
| Raman tensor *(unique axis b)** | $\begin{pmatrix} \tilde{a} & 0 & \tilde{d} \\ 0 & \tilde{b} & 0 \\ \tilde{d} & 0 & \tilde{c} \end{pmatrix}$ | $\begin{pmatrix} 0 & \tilde{f} & 0 \\ \tilde{f} & 0 & \tilde{e} \\ 0 & \tilde{e} & 0 \end{pmatrix}$ | c, b, a, β ≠ 90° |
| Raman tensor *(unique axis c)** | $\begin{pmatrix} \tilde{a} & \tilde{d} & 0 \\ \tilde{d} & \tilde{b} & 0 \\ 0 & 0 & \tilde{c} \end{pmatrix}$ | $\begin{pmatrix} 0 & 0 & \tilde{e} \\ 0 & 0 & \tilde{f} \\ \tilde{e} & \tilde{f} & 0 \end{pmatrix}$ | c, b, a, γ ≠ 90° |

(* from Bilbao Crystallographic Server[1])

**Table S2.** Measured and calculated Raman peak frequencies in CVT-grown $AgVP_2Se_6$ along with their corresponding mode symmetries. Our work is compared with Chen *et al.*'s report at room temperature (RT).[2]

| This work (5 K) | | | | Ref [2] (RT) | |
|---|---|---|---|---|---|
| $\omega_{exp,5\ K}$(cm$^{-1}$) | Symmetry (Exp) | $\omega_{DFT}$(cm$^{-1}$) | Symmetry (DFT) | $\omega_{exp}$(cm$^{-1}$) | Symmetry (Exp) |
| 39.6 | A | 35.5 | B | | |
| 45.7 | B | 46.5 | A | 43 | B |
| 56.8 | A | 55.5 | B | | |
| 70.9 | A | 70.1 | B | | |
| 77.3 | A | 81.5 | A | 76 | B |
| 82.5 | B | 101 | B | | |
| 112.1 | A | 110.1 | A | 113 | B |
| | | 117.8 | B | | |
| 119.2 | A | 118 | A | | |
| 124.3 | A | 126.5 | A | | |
| 129.1 | B | 126.8 | B | | |
| 140.5 | A | 136.4 | B | | |
| 146.3 | B | 140.5 | B | 146 | B |
| 148.0 | A | 141.9 | A | | |
| 161.3 | B | 158.5 | A | | |
| 163.2 | B | 159.8 | B | 162 | B |
| 164.6 | A | 160.1 | A | | |
| 210.6 | A | 197.8 | A | 209 | B |
| 231.6 | A | 225.7 | A | | |
| 237.1 | A | | | | |
| 243.9 | B | 236.2 | B | | |
| 262.4 | A | 259.2 | B | | |
| 320.3 | A | 314.8 | B | | |
| 428.3 | A | | | | |
| 432.2 | B | 404.6 | B | 430 | B |
| 434.9 | A | 419.4 | A | | |
| 436.3 | B | | | | |
| 444.1 | A | 424.6 | A | | |
| 446.1 | B | 435.9 | B | | |
| 452.5 | A | | | | |
| 515.6 | A | 479.3 | A | | |

**Table S3.** Atomic displacements for the Raman modes calculated by DFT. Phonon eigenvectors for the 27 Raman active peaks in $AgVP_2Se_6$. The Ag, V, P, and Se atoms are labeled in silver, yellow, green, and purple, respectively.

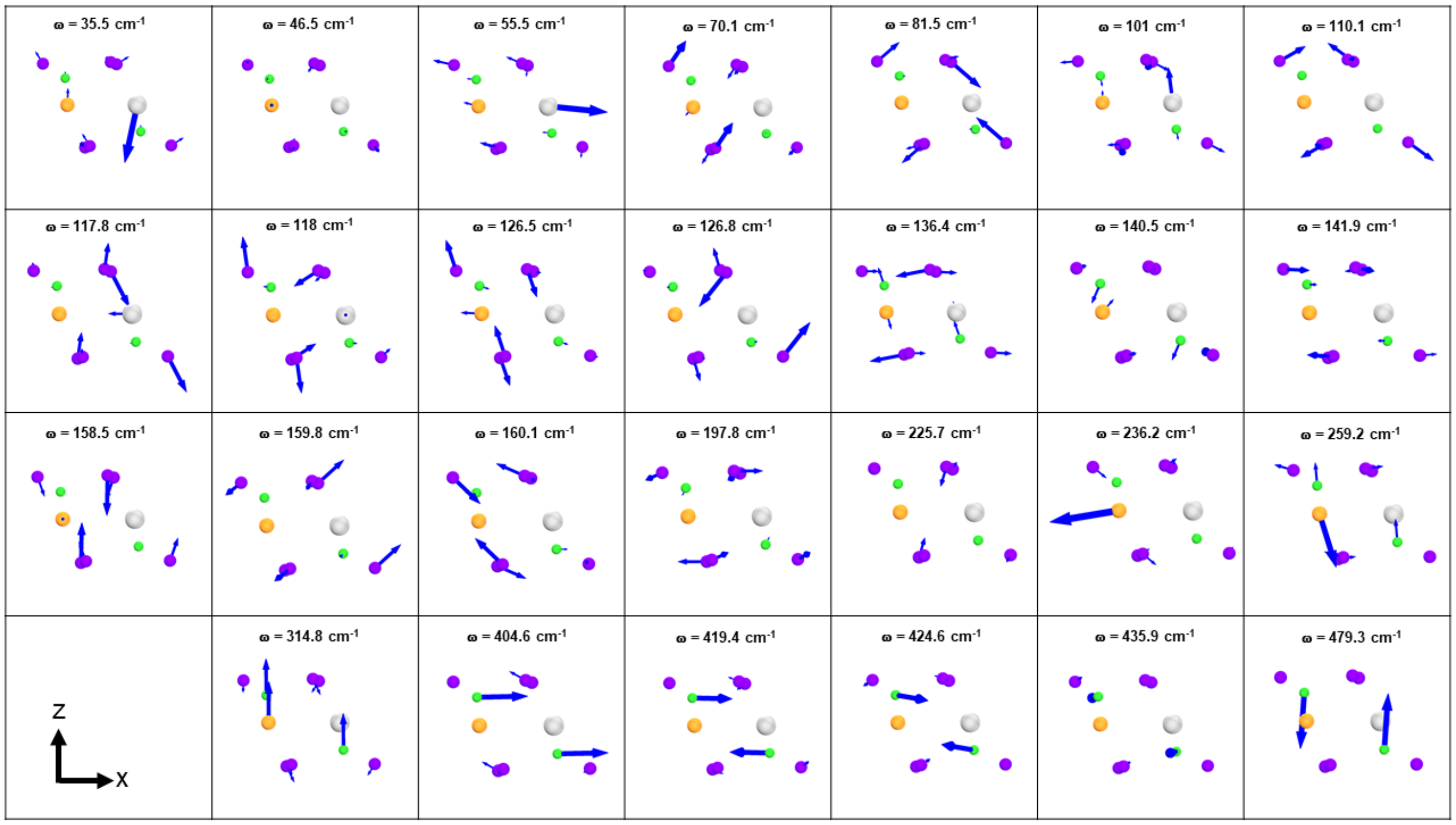

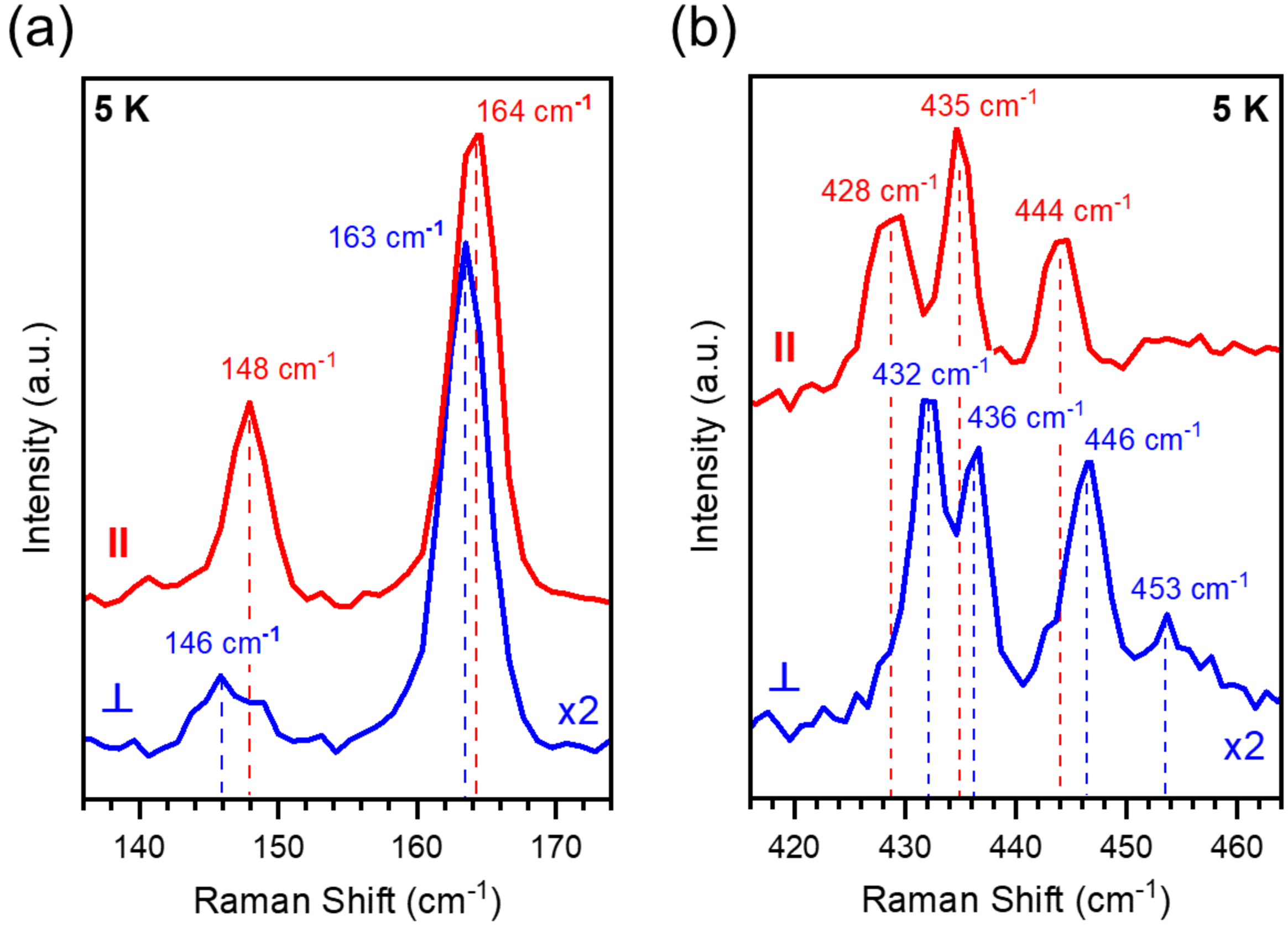

**Figure S1.** Polarized Raman spectra when the polarization direction of the incident polarized light $\hat{\boldsymbol{e}}_i$ perpendicular to the crystallographic ***a***-axis shows several hybridized modes at (a) 146 $cm^{-1}$ - 148 $cm^{-1}$, 163 $cm^{-1}$ - 164 $cm^{-1}$, (b) 435 $cm^{-1}$ - 436 $cm^{-1}$, and 444 $cm^{-1}$ - 446 $cm^{-1}$.

Based on the polarization equations for A and B modes, $I_A^{\perp} \propto \sin^2(\phi+\phi_0)\cos^2(\phi+\phi_0)\,(a^2+b^2-2ab\cos\varphi_{ab})$ and $I_B^{\parallel} \propto f^2\sin^2 2(\phi+\phi_0)$, when $\phi+\phi_0 = 90°$ or the incident polarized light $\hat{\boldsymbol{e}}_i$ perpendicular to the crystallographic ***a***-axis, $I_A^{\perp} = I_B^{\parallel} = 0$. Contrasting the co- and cross- polarization Raman spectra allows us to identify the adjacent modes at 146 $cm^{-1}$ (B), 148 $cm^{-1}$ (A), 163 $cm^{-1}$ (B), 164 $cm^{-1}$ (B), 432 $cm^{-1}$(B), 435 $cm^{-1}$(A), 436 $cm^{-1}$(B), 444 $cm^{-1}$ (A), and 446 $cm^{-1}$ (B).

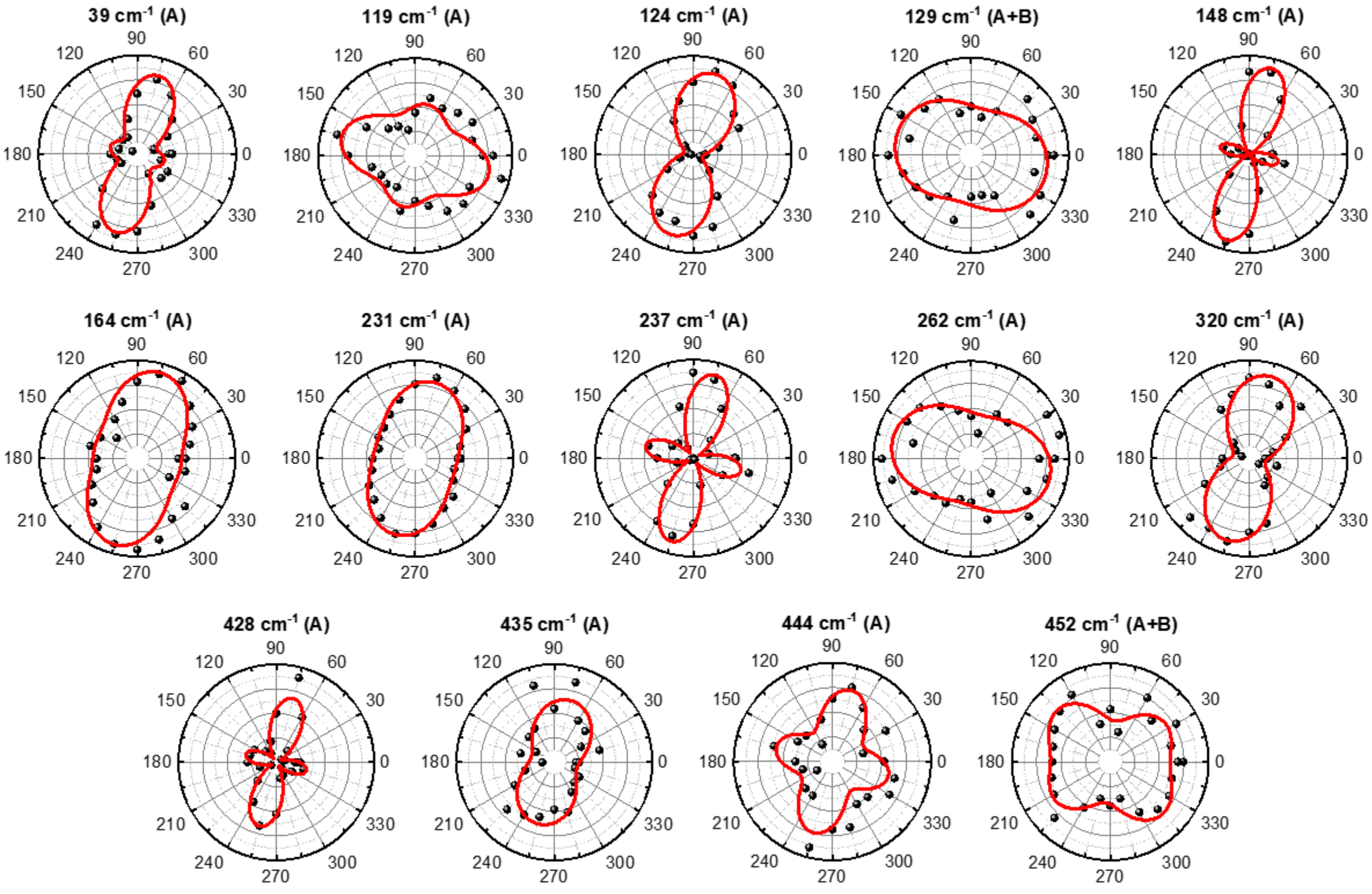

**Figure S2.** Angular dependence of Raman integrated intensities of A-modes taken in co-parallel (∥) polarization configuration with 532 nm laser at 5 K. Solid lines are the best fits to the data using eq 4 in the main text for $I_A^{\parallel}$. The combination of eqs 4 and 6 for $I_A^{\parallel} + I_B^{\parallel}$ is required to fit some hybrid modes such as 129 cm$^{-1}$ and 452 cm$^{-1}$.

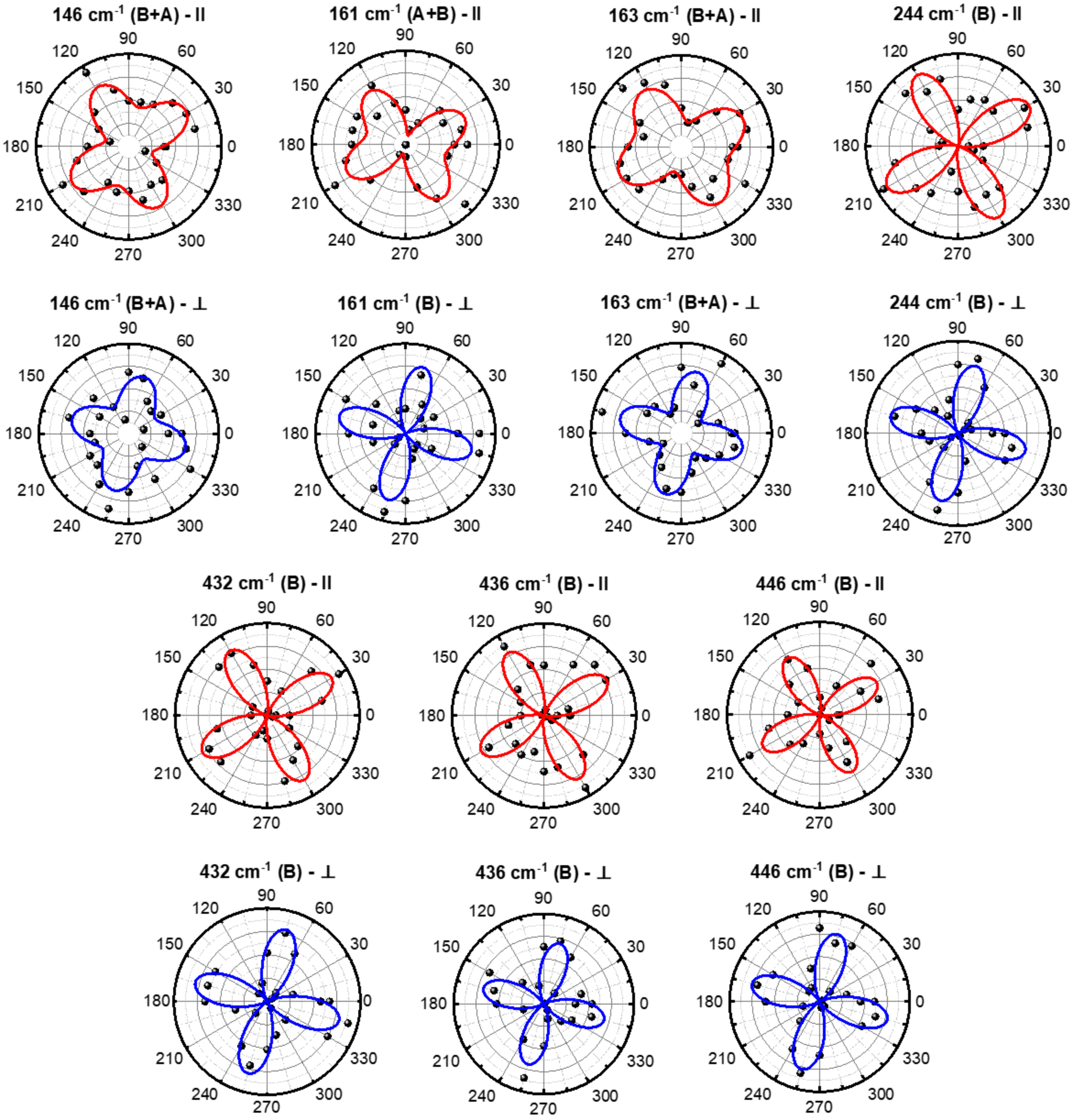

**Figure S3.** Angular dependence of Raman integrated intensities of B-modes taken with 532 nm laser at 5 K. Polar graphs of parallel (red) and cross (blue) polarization configurations for all B modes. Solid lines are the best fits to the data using eqs 6 and 7 in the main text for $I_B^{\parallel}$ and $I_B^{\perp}$, respectively. The combined equations for $I_B^{\parallel} + I_A^{\parallel}$ and $I_B^{\perp} + I_A^{\perp}$ are required to fit some hybrid modes such as 146 cm$^{-1}$, 161 cm$^{-1}$ and 163 cm$^{-1}$.

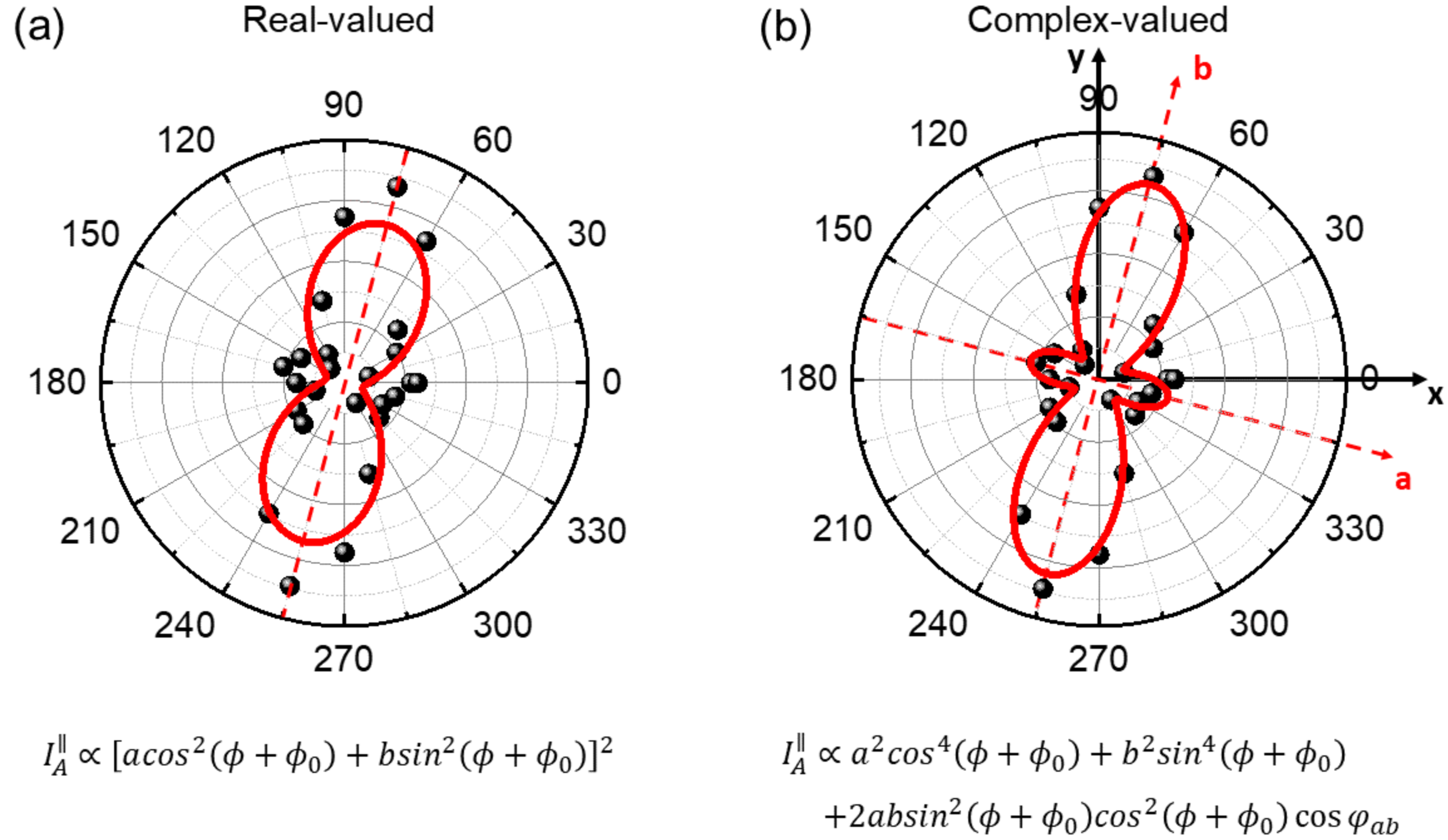

**Figure S4.** The angle dependent Raman intensity of the 56 cm$^{-1}$-mode is fitted using equations derived from real-valued (a) and complex-valued (b) Raman tensors. The former one fails to reproduce the secondary maxima at 165° and 345° in the polar plot. Red dash line/arrows represent the crystallographic ***a*** and ***b*** axes which make an offset angle of approximately 15° with respect to the lab frame ***x*** and ***y*** axes, respectively.

**Table S4.** Tensor components for A modes retrieved from ARRS data in parallel configuration taken at 5 K

| $\omega_{exp}$(cm$^{-1}$) | fitting function $I_A^{\parallel}$ | $a/b$ | $\varphi_{ab}$ (°) | $f/a$ |
|---|---|---|---|---|
| 39.6 | A | 0.59 | 91.3 $\pm$ 8.9 | |
| 56.8 | A | 0.59 | 121.2 $\pm$ 6.6 | |
| 77.3 | A | 0.55 | 63 $\pm$ 8.5 | |
| 112.1 | A | 0.81 | 0 | |
| 119.2 | A | 1.2 | 63.9 $\pm$ 11.8 | |
| 124.3 | A | 0.33 | 0 | |
| 129 | A+B | 1.27 | 0 | 0.20 |
| 148.0 | A | 0.58 | 179.7 $\pm$ 49 | |
| 164.6 | A | 0.70 | 41.3 $\pm$ 13.9 | |
| 210.6 | A | 0.58 | 0 | |
| 231.6 | A | 0.73 | 30.3 $\pm$ 8.4 | |
| 237.1 | A | 0.76 | 143.7 $\pm$ 9.3 | |
| 262.4 | A | 1.34 | 0 | |
| 320.3 | A | 0.44 | 0 | |
| 428.3 | A | 0.69 | 141.5 $\pm$ 10.1 | |
| 434.9 | A | 0.61 | 0 | |
| 444.1 | A | 0.88 | 86.2 $\pm$ 10.4 | |
| 452.5 | A+B | 1.22 | 0 | 0.69 |
| 515.6 | A+B | 0.99 | 0 | 0.54 |

## Angle-resolved Raman polarization of 515 $cm^{-1}$ ($AgVP_2Se_6$) and 524 $cm^{-1}$ (Si) modes

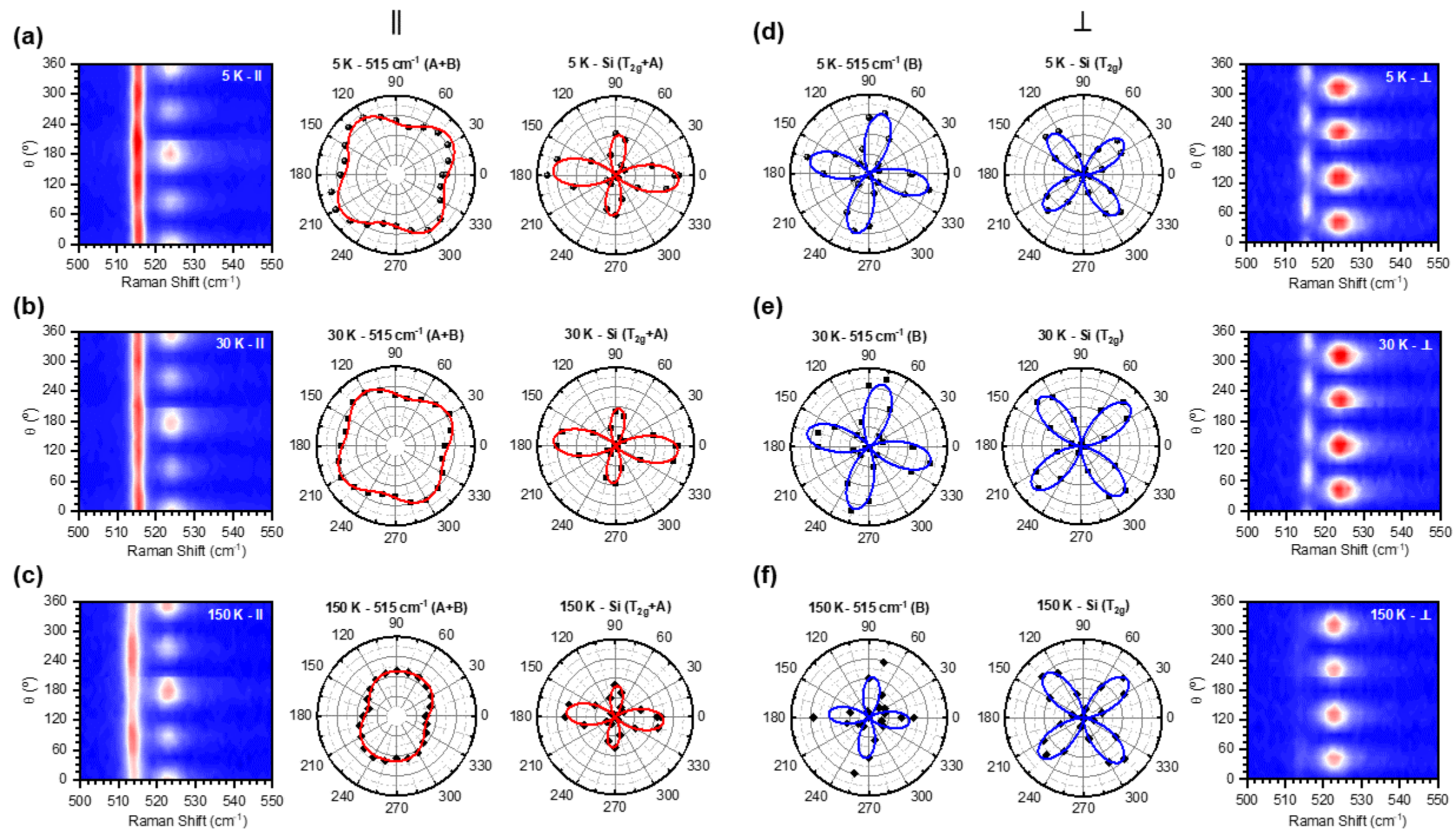

**Figure S5. Anomalous angle-resolved Raman polarization of $AgVP_2Se_6$ and $SiO_2$/Si substrate**. Each panel includes, from left to right, the false-color map and the angle-dependent integrated intensity plots of 515 $cm^{-1}$ ($AgVP_2Se_6$) and 524 $cm^{-1}$ (Si) modes in parallel (a-c) and cross (d-f) configurations. Data is taken at three temperatures: 5 K (a, d), 30 K (b, e) and 150 K (c, f).

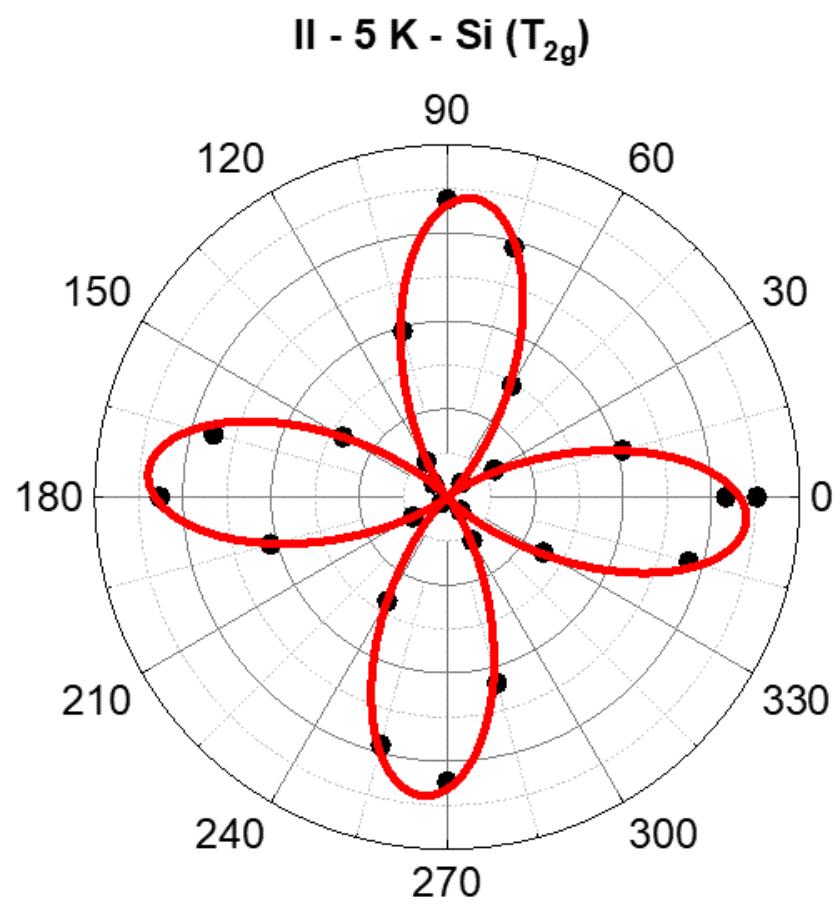

**Figure S6.** Angle-dependent Raman intensity of $T_{2g}$ mode in the bare $SiO_2$/Si substrate without the $AgVP_2Se_6$ overlayer. Data is measured in the parallel polarization configuration at 5 K.

**Table S5.** Fitting parameters for the 515 cm$^{-1}$ ($AgVP_2Se_6$) and 524cm$^{-1}$ (Si) polar plots in parallel (∥) and cross (⊥) configurations. Parameters for bare $SiO_2$/Si substrate without flake are also given for comparison.

| | | 5 K | | 30 K | | 150 K | |
|---|---|---|---|---|---|---|---|
| | | ∥ | ⊥ | ∥ | ⊥ | ∥ | ⊥ |
| **515 cm$^{-1}$** | $a$ | 41.78 ± 0.67 | | 42.08 ± 0.24 | | 32.46 ± 0.34 | |
| | $b$ | 41.87 ± 0.71 | | 41.41 ± 0.2 | | 39.78 ± 0.30 | |
| | $f$ | 23.78 ± 1.55 | 21.47 ± 0.24 | 21.10 ± 0.65 | 22.05 ± 0.28 | 12.52 ± 1.18 | 11.17 ± 0.13 |
| | $\phi_0$ | 10.9° ± 1.9° | 13.4° ± 0.9° | 12.1° ± 0.9° | 13.3° ± 0.6° | 11.9° ± 1.4° | 5.6° ± 2.5° |
| **Si with flake** | $a$ | 26.71 ± 5.25 | | 36.77 ± 1.99 | | 31.21 ± 1.48 | |
| | $b$ | 0 ± 5.62 | | 0 ± 5.39 | | 0 ± 4.12 | |
| | $\phi_0^A$ | 4.5° ± 4.4° | | 5.9° ± 3.2° | | 8.2° ± 2.8° | |
| | $t$ | 47.55 ± 2.34 | 51.07 ± 0.29 | 45.26 ± 1.32 | 47.55 ± 0.23 | 41.40 ± 0.91 | 45.44 ± 0.20 |
| | $\phi_0^{Si}$ | 49.4° ± 1.2° | 49.3° ± 0.3° | 49.1° ± 1.2° | 49.5° ± 0.2° | 48.8° ± 0.9° | 48.5° ± 0.3° |
| **bare** | $t$ | 33.26 ± 0.34 | | | | | |
| | $\phi_0^{Si}$ | 49.6° ± 0.3° | | | | | |

Here, the near-isotropic, four-fold symmetric polarization pattern of 515 cm$^{-1}$ mode in the parallel (∥) configuration (**Figures S5a-c**) is fitted using the combined function $I_A^{\parallel} + I_B^{\parallel} \propto a^2 \cos^4(\phi + \phi_0) + b^2 \sin^4(\phi + \phi_0) + 2ab \sin^2(\phi + \phi_0) \cos^2(\phi + \phi_0) \cos \varphi_{ab} + f \sin^2 2(\phi + \phi_0)$. However, in the cross polarization at all three temperatures, the polar plots of this peak can be sufficiently described by only B-symmetry function, $I_B^{\perp} \propto f^2 \cos^2 2(\phi + \phi_0)$, as shown in **Figures S5d-f**. Extracted fitting parameters are presented in **Table S5** below. The offset angle $\phi_0$ varies in the range of 10° - 14°, in close agreement with the predetermined value 15° mentioned earlier in the main text.

On the other hand, the Si (100) substrate is characterized by its $T_{2g}$ mode (524 cm$^{-1}$) at 5 K. In the backscattering geometry $-\boldsymbol{z}(\hat{\boldsymbol{e}}_i\hat{\boldsymbol{e}}_s)\boldsymbol{z}$, where the incident and scattered light propagate along the $\boldsymbol{z}$-axis, only the phonon response of the $T_{2g(z)}$ degenerate mode survives[3], and its tensor is given as

$$R_{T_{2g(z)}} = \begin{pmatrix} 0 & t & 0 \\ t & 0 & 0 \\ 0 & 0 & 0 \end{pmatrix} \quad (1)$$

Thus, we calculate the ideal Raman intensity for $T_{2g}$ phonon from Si substrate as

$$I_{T_{2g}}^{\parallel} \propto t^2 \sin^2 2\left(\phi + \phi_0^{Si}\right) \quad (2)$$

$$I_{T_{2g}}^{\perp} \propto t^2 \cos^2 2\left(\phi + \phi_0^{Si}\right) \quad (3)$$

where $\phi_0^{Si}$ indicates the relative incident polarization with respect to the in-plane crystallographic axis of Si substrate, and $\phi$ is the rotation angle of $\hat{e}_i$ by sweeping the half wave plate. These equations are analogous to the polarization equations of *B*-phonon; hence, one would expect the similar 4-fold symmetry of Si (100) polar plots, as shown in **Figure S6** of the bare substrate. Extracted angle $\phi_0^{Si} \approx 49°$ **(Table S5)** implies the fixed orientation of the Si substrate. **Figure S5a** shows an anisotropic polarization pattern of the flake-covered $SiO_2$/Si where the sole $T_{2g}$-function fails to fit the angle-dependent Raman intensity of 524 cm$^{-1}$ (Si) peak in the parallel (∥) configuration. In fact, the resulting two-fold symmetry of the Si mode indicates possible influence of the A-mode behavior, suggesting the alternative fitting function as

$$I_{T_{2g}}^{\parallel} + I_A^{\parallel} \propto t^2 \sin^2 2\left(\phi + \phi_0^{Si}\right) + a^2 \cos^4\left(\phi + \phi_0\right) + b^2 \sin^4\left(\phi + \phi_0\right)$$
$$+2ab \sin^2\left(\phi + \phi_0\right) \cos^2\left(\phi + \phi_0\right) \cos \varphi_{ab} \quad (4)$$

To explain why this admixture is confined to the parallel configuration and carries a phase offset matching neither the Si nor the $AgVP_2Se_6$ axis (**Table S5**), we model $AgVP_2Se_6$ as a linear dichroic (anisotropic-transmission) optical element using standard Jones calculus.[4] In its own principal-axis frame, such a medium has the diagonal Jones matrix

$$J = \begin{pmatrix} t_a & 0 \\ 0 & t_b \end{pmatrix}$$

where $t_a$ and $t_b$ are the (real) field-transmission amplitudes along the ***a***- and ***b***-axes, generally unequal. For example, angle-resolved polarized absorption spectroscopy and first-principles calculations on the sulfide analogue $AgVP_2S_6$ report pronounced in-plane anisotropy in both the real and imaginary parts of the dielectric function, i.e., simultaneous birefringence and linear dichroism along the ***a***- and ***b***-axes.[5] For an incident field linearly polarized at angle $(\phi + \phi_0)$ relative to the ***a***-axis, the Jones vector, normalized to unit incident intensity, is $\boldsymbol{E_{in}} = [\cos(\phi + \phi_0), \sin(\phi + \phi_0)]^T$, and the transmitted field is $\boldsymbol{E_{out}} = J.\boldsymbol{E_{in}} = [t_a \cos(\phi + \phi_0),\ t_b \sin(\phi + \phi_0)]^T$. Each component is attenuated by its own axis's amplitude transmittance, so $\boldsymbol{E_{out}}$ is generally no longer parallel to $\boldsymbol{E_{in}}$ unless $t_a = t_b$. The total transmitted intensity is given as

$$T(\phi) = |\boldsymbol{E_{out}}|^2 = t_a^2 \cos^2(\phi + \phi_0) + t_b^2 \sin^2(\phi + \phi_0) \quad (5)$$

In this backscattering geometry, light travels through the flake twice: once on the incident path down to the substrate, and once on the collected path as the Si-scattered light returns through the flake to the analyzer. Each pass is an independent power-transmission event governed by eq 5, evaluated at whatever polarization angle is relevant to that pass. In the parallel configuration, both the incident polarizer and the collection analyzer are set to the same angle $\phi$, so both passes see the same transmission factor $T(\phi)$, and the net attenuation is the product $T(\phi) \cdot T(\phi) = [T(\phi)]^2$. In the cross configuration, the incident and collected polarizations are orthogonal, so the incident path sees $T(\phi)$ while the collected path sees $T(\phi + 90°)$ instead, giving the product $T(\phi) \cdot T(\phi + 90°)$. Combining this double-pass transmission with the Raman-intensity projection formalism used throughout this work gives the substrate signal in each configuration. In the parallel configuration, both the incident and collected light sample this same anisotropic transmission

$$I_{Si}^{\parallel}(\phi) \propto I_{T_{2g}}^{\parallel}(\phi) \cdot [T(\phi)]^2 \quad (6)$$

while in the cross configuration

$$I_{Si}^{\perp}(\phi) \propto I_{T_{2g}}^{\perp}(\phi) \cdot T(\phi) \cdot T(\phi + 90°) \quad (7)$$

Equations 6 and 7 use $T(\phi)$ as a scalar transmission factor on each one-way pass, rather than an exact two-pass calculation. Although it would not track how the light's polarization angle rotates after the first pass, the scalar factorization used here matches the level of approximation used in the established substrate/overlayer interference framework for isotropic 2D materials on $SiO_2$/Si.[6] By rewriting $I_{T_{2g}}^{\parallel}(\phi) \propto \left(\frac{t^2}{2}\right)\left[1-\cos\left(4\phi+4\phi_0^{Si}\right)\right]$, $I_{T_{2g}}^{\perp}(\phi) \propto \left(\frac{t^2}{2}\right)\left[1+\cos\left(4\phi+4\phi_0^{Si}\right)\right]$ and $T(\phi)=T_0+\Delta T\cos(2\phi+2\phi_0)$ with $T_0=(t_a^2+t_b^2)/2$ and $\Delta T=(t_a^2-t_b^2)/2$, eqs 6 and 7 become

$$I_{Si}^{\parallel}(\phi) \propto \left(\frac{t^2}{2}\right)\left[1-\cos\left(4\phi+4\phi_0^{Si}\right)\right][T_0+\Delta T\cos(2\phi+2\phi_0)]^2 \qquad (8)$$

$$I_{Si}^{\perp}(\phi) \propto \left(\frac{t^2}{2}\right)\left[1+\cos\left(4\phi+4\phi_0^{Si}\right)\right][T_0^2-\Delta T^2\cos^2(2\phi+2\phi_0)] \qquad (9)$$

**Parallel configuration:** If $t_a=t_b$, $T(\phi)$ would be a constant, and eq 6 would simply rescale the Si tensor's intrinsic four-lobed pattern without distorting it. When $t_a \neq t_b$, $T(\phi)$ instead oscillates with $\phi$, peaking when the polarization aligns with the higher-transmission axis and dipping when it aligns with the lower-transmission axis. Multiplying this angle-dependent factor into the Si tensor's own four-lobed pattern in eq 6 results in unequal maximum intensity of adjacent lobes. This matches our observation of the Si polar plot distortion in (‖) and its offset angles in **Table S5**, which reflects neither axis alone.

**Cross configuration:** The oscillating term of $T(\phi+90°)$ flips sign from $T(\phi)$. Whenever the incident pass is boosted by aligning with the high-transmission axis, the analyzer pass at $\phi+90°$ is simultaneously cut by aligning with the low-transmission axis, and vice versa as $\phi$ sweeps around. This anti-correlation cancels the lobe-unbalancing term in the product $T(\phi)\cdot T(\phi+90°)$, leaving only a symmetric modulation with the same four-fold periodicity as the Si tensor itself. This is why in the cross-configuration, Si's polar plot recovers an equal-lobed four-fold pattern (**Figure S5d-f, Table S5**).

The approach from standard dichroic-media optics[4] is motivated by the substrate/overlayer interference effect established for isotropic 2D materials on $SiO_2$/Si[6] and by recent multilayer transfer-matrix treatments of angle-resolved polarized Raman intensity that incorporate polarization-dependent interfacial optics for anisotropic flakes.[7] Equations 6 and 7 should accordingly be considered as a proposed model to describe a substrate signal beneath an anisotropic overlayer, which has not been addressed in the literature.

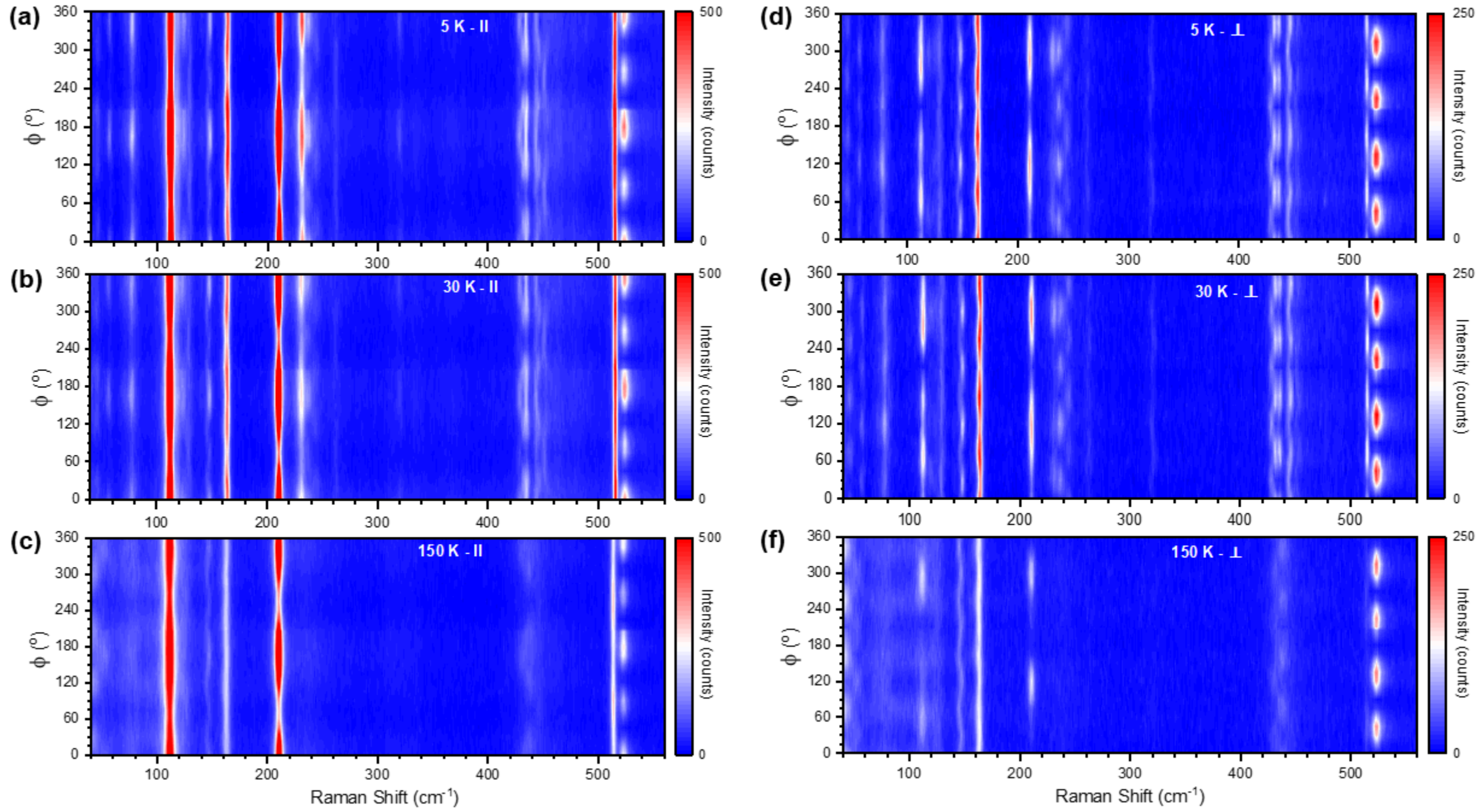

**Figure S7.** Angle-resolved Raman polar maps of the CVT-grown $AgVP_2Se_6$ flake in co- and cross- polarization configurations at (a) 5 K, (b) 30 K, and (c) 150 K.